\documentclass[reprint, showpacs,preprintnumbers,onecolumn,amsmath,amssymb,aps,10pt,pre,floatfix,longbibliography]{revtex4-1}

\usepackage{geometry}
\usepackage{multirow}
\usepackage{color}
\usepackage{graphicx}
\usepackage{dcolumn}
\usepackage{bm}
\usepackage{longtable}
\usepackage{float}
\usepackage{subfigure}
\usepackage{epstopdf}
\usepackage{soul}
\usepackage[normalem]{ulem}

\usepackage[pdftex,bookmarks,colorlinks]{hyperref}
\pdfoutput=1
\usepackage{amsmath}
\usepackage{amssymb}

\newcommand{\lra}[1]{\langle #1 \rangle }

\newcommand{\gv}[1]{\ensuremath{\mbox{\boldmath$ #1 $}}} 
\newcommand{\grad}[1]{\gv{\nabla} #1} 
\renewcommand{\div}[1]{\gv{\nabla} \cdot #1} 
\newcommand{\erf}[1]{\text{ erf}\left(#1\right)} 

\begin{document}

\title{Basal melting driven by turbulent thermal convection}

\author{Babak Rabbanipour Esfahani}\email[]{babak.rabbanipour@polytech-lille.fr}
\author{Silvia C. Hirata}
\author{Stefano Berti}
\author{Enrico Calzavarini}

\affiliation{Univ. Lille, Unit\'e de M\'ecanique de Lille, UML EA 7512, F-59000 Lille, France}

\date{\today}

\begin{abstract}
Melting and, conversely, solidification processes in the presence of convection are key to many geophysical problems.  
An essential question related to these phenomena concerns the estimation of the (time-evolving) melting rate, which is tightly 
connected to the turbulent convective dynamics in the bulk of the melt fluid and the heat transfer at the liquid-solid interface. 
In this work, we consider a convective-melting model, constructed as a generalization of the 
Rayleigh-B\'enard system, accounting for the basal melting of a solid. As the change of phase proceeds, a fluid layer grows at the 
heated bottom of the system and eventually reaches a turbulent convection state. By means of extensive Lattice-Boltzmann numerical 
simulations employing an enthalpy formulation of the governing equations, we explore the model dynamics in two and 
three-dimensional configurations. The focus of the analysis is on the scaling of global quantities like the heat flux and 
the kinetic energy with the Rayleigh number, as well as on the interface morphology and the effects of space dimensionality. 
Independently of dimensionality, we find that the convective-melting system behavior shares strong resemblances with that of the 
Rayleigh-B\'enard one, and that the heat flux is only weakly enhanced with respect to that case. Such similarities are understood, 
at least to some extent, considering the resulting slow motion of the melting front (with respect to the turbulent fluid velocity fluctuations) 
and its generally little roughness (compared to the height of the fluid layer). Varying the Stefan number, accounting for the 
thermodynamical properties of the material, also seems to have only a mild effect, which implies the possibility to 
extrapolate results in numerically delicate low-Stefan setups from more convenient high-Stefan ones. 
Finally, we discuss the implications of our findings for the geophysically relevant problem of modeling Arctic ice melt ponds. 
\end{abstract}

\maketitle

\section{Introduction \label{sec:intro}}
Melting and solidification coupled with convective flows are fundamental processes in the geophysical context. Convective melting (CM) is thought to have played a major role in Earth's mantle formation \cite{solomatov} and is commonly observed in magma chambers \cite{brandeis1986interaction,brandeis1989convective}, lava lakes \cite{davaille1993thermal} or melt-ice lakes \cite{Polashenski2012,Polashenski2017}. 
All these systems are characterized by the presence of unsteady, chaotic and often turbulent flows. 
Turbulence arising from natural convection in fixed-shape domains like the Rayleigh-B\'enard (RB) system 
has been studied in depth through laboratory experiments, as well as theoretical and numerical investigations 
\cite{chandrasekhar2013hydrodynamic, Chilla, ahlers2009heat}. 
Much less attention has been instead directed to the problem of coupled turbulent natural convection and phase change, 
particularly in the case of basal (as opposed to lateral) heating. A key difference between the CM and RB dynamics 
pertains to the role of time. While the RB system can be considered stationary in a statistical sense, the CM system is intrinsically 
non-stationary due to the phase change occurring at boundaries and leading to a continuously time varying fluid domain.
An important question related to convective melting processes in all their generality is the prediction of the evolution 
of the melting rate, which is connected to the heat-flux dynamics determined by the flow in the system. To what extent the knowledge 
acquired on turbulent, i.e. high-Rayleigh number, natural convection can be 
exploited to understand convective melting is at the moment an open problem and a central question in this paper. 
A second open issue that deserves attention concerns the characterization, at least in a statistical sense, of the shape 
(often termed topography in geophysical applications) of the solid-to-liquid phase-change interface resulting from 
the turbulent convective transport of heat.

The study reported in this article was motivated by the ongoing research on the dynamics of ice melt ponds that form 
during the summer season in the Arctic \cite{Polashenski2012}. 
Such ponds absorb heat from a source situated on their top side.  
Heat absorption is partly due to the contact with warmer air ($\sim2^o C$) and partly to a volumetric 
contribution from solar radiation. At the bottom of the pond the melt water is instead in contact with ice ($\leq 0^o C$). 
Since the typical temperatures involved are definitely lower than $4^o C$ \cite{taylor2004model}, 
the density of the fresh water contained in ponds increases 
with increasing temperature. Such fluid layers then result to be in a dynamically unstable state and display natural convection 
coupled to a phase-change process on the bottom side.
Ice melt ponds are  known to have an important role in the global climate dynamics because they strongly affect the 
effective albedo (i.e. the ratio of reflected over incoming solar radiation) in polar regions. 
Because water is a good absorber of electro-magnetic radiation, the low albedo of ponds, compared to that 
of snow or sea ice, causes them to preferentially absorb heat, which further affects the bottom side sea-ice melting 
through a positive feedback mechanism \cite{Deser2000}. 
A better understanding of the small-scale (few to some tens of centimeters) mechanisms controlling the convective 
heat transfer in ponds is necessary in order to provide useful guidelines for parameterizations in large-scale 
ice models \cite{flocco2007continuum,scott2010,vancoppenolle2009simulating}.
The rate of melting in water ponds depends on a variety of factors including the temperature of the air, 
the effect of wind draft, the residual salinity of water in the ponds, and the intensity of the buoyancy force 
leading to convection. Among these factors, convection plays a prominent role, 
as it enhances water mixing and increases the total intake of energy into the system. The flow in ponds is generally 
turbulent \cite{taylor2004model}, with realistic values of the Rayleigh number in the range $10^6$-$10^9$. 
In addition, the topography of the bottom surface of a melt pond can affect the flow, by creating flow patterns and 
coherent thermal structures that can differ from those occurring between flat plates or plates with prescribed roughness \cite{Chilla}. 
Because the absorption of solar radiation is a nonlinear function of the water layer depth, the evolving bottom topography of ponds is also a key parameter for precise estimations of the pond albedo \cite{podgorny1996}.

In the present work, we investigate the behavior of a model system in which a pure substance initially in the solid state 
is progressively melted by a horizontal heat source. The melt fluid layer is thermally unstable and quickly develops convective 
motion of progressively higher intensity as the depth of the melt layer increases. This simple realization of 
basal-heating driven convective melting allows thorough analyses of the dependencies 
of global flow observables, such as the total heat flux and the total kinetic energy, 
on the varying melt fluid layer depth.
It also allows to reveal the possible links between the flow and the phase-change interface shaped by it.

The remainder of this paper is organized as follows. In section \ref{sec:model} we introduce 
the adopted model system together with its evolution equations. A discussion about the global heat flux budget 
with additional dimensional arguments for the heat flux scaling behavior in different flow regimes is presented 
in Sec. \ref{sec:math}. 
Section \ref{sec:methods} concisely presents the numerical simulations, which are implemented via 
a Lattice-Boltzmann method capable to accurately describe both the turbulent convective dynamics of the melt water and 
the solid-to-liquid phase change.
The results of simulations in two and three dimensions (2D, 3D, respectively) are presented and discussed in Sec. \ref{sec:results}. 
To interpret and rationalize the observed trends in the scaling of global quantities, such as the Nusselt and Reynolds numbers, 
we specialize the discussion to the effects of space dimensionality, Sec. \ref{sec:effect2d} and Sec. \ref{sec:effect3d}, 
and we analyze the morphology of the melting front in section \ref{sec:interface}. 
The effects of varying the Stefan number on the melting rate is studied in Sec. \ref{sec:stdependency}. 
Final discussions and conclusions are given in Sec. \ref{sec:concl}.

\section{The convective melting system with basal heating \label{sec:model}}
\begin{figure}[!htb]
\centering
\includegraphics[width=0.95\linewidth]{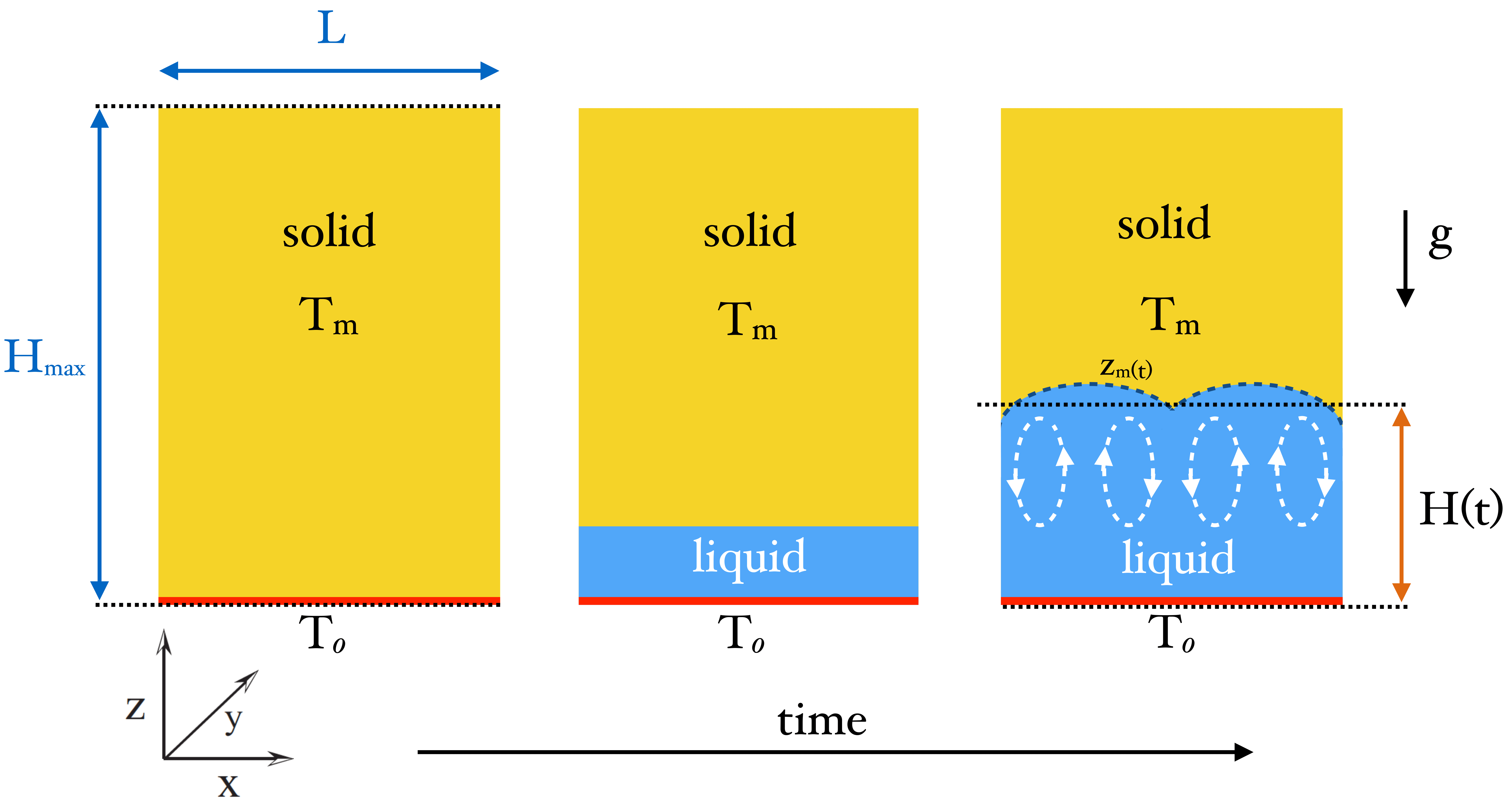}
\caption{Schematic view (2D cut) of the melting system. This initially consists of a pure solid at the 
melting temperature ($T_m$). The bottom temperature is kept fixed at a value $T_0>T_m$, allowing the formation of a liquid phase. 
As time progresses the bottom fluid layer becomes deeper and convection can develop in it. The local height of the liquid-solid interface, 
measured from the bottom, is denoted $z_m$; its average over horizontal ($x,y$) coordinates is $H$.} 
\label{fig:fig01}
\end{figure}

The model system considered in this study consists of a  solid layer of a pure substance of thickness $H_{max}$  initially at a constant temperature, $T_m$, equal to the phase change (melting) temperature.  
At  time $t>0$ the bottom boundary of the solid is heated at a constant temperature $T_0 > T_m$ and a melted fluid layer 
begins to grow from below with the liquid-solid interface advancing in the direction opposite to gravity.  
The density of the fluid is assumed to be a decreasing function of temperature, therefore the bottom heating produces an 
unstable stratification of the fluid layer. 
A cartoon of the model system is shown in Fig.~\ref{fig:fig01}.\\  
We shall note that our model system is dynamically equivalent to the setting mentioned earlier of an  Arctic melt pond, 
although it is an upside-down representation of it. Indeed, in melt ponds, heating occurs at the top rather than 
at the bottom but warmer water parcels are negatively (instead of positively) buoyant. 
Notice, however, that for simplicity we neglect the distributed thermal forcing due to solar radiation and  wind-induced shear at the hot boundary (the air-water interface for real ponds).

\subsection{Equations of motion \label{sec:eqs_motion}}
Under the assumption that the temperature differences occurring in the system are small enough for 
the Boussinesq approximation to hold, the governing equations in the melt layer, are:
\begin{eqnarray}
\rho_0 \left(\ \partial_t \bm{u}+\bm{u}\cdot\grad{\bm{u}}\ \right)&=& -\grad{p}+\mu\bm{\nabla}^2\bm{u}+
\rho_l\bm{g},\label{eq:vel}\\
\div{\bm{u}}&=&0, \label{eq:nodivergence}\\
\rho_l&=&\rho_0 \left(1-\beta(T-T_{0})\right), \label{eq:stateeq}\\
\partial_t T+\bm{u}\cdot\grad{T}&=&\kappa\bm{\nabla}^2 T, \label{eq:T} 
\end{eqnarray}
where $\bm{u}(\bm{x},t)$,  $p(\bm{x},t)$ and  $T(\bm{x},t)$ respectively are the fluid velocity, 
pressure and temperature fields; $\mu$ is the dynamic viscosity and $\bm{g}$ is gravity acceleration. 
The fluid density $\rho_l$ is assumed to linearly depend on temperature, with $\rho_0$ being the reference 
density at temperature $T_0$ and $\beta$ the thermal expansion coefficient; $\kappa$ indicates the thermal diffusivity. 
Note that the flow is incompressible, as a consequence of the Boussinesq approximation.

The boundary conditions associated to the above set of equations are isothermal and no-slip for temperature 
and velocity, respectively, at the bottom wall, periodic at lateral boundaries, and  
no slip and melting (i.e. Stefan condition for a solid at melting temperature \cite{alexiades1992mathematical}) 
at the phase-change interface. 
The conditions at vertical boundaries then read:
\begin{eqnarray}
T|_{\bm{x}=(x,y,0)} & = & T_0 \quad \forall \, (x,y) \in [0,L]^2, \label{eq:bc1T}\\
\bm{u}|_{\bm{x}=(x,y,0)} & = & \bm{0} \quad \forall \, (x,y) \in [0,L]^2,\\
- \kappa\ \bm{\nabla}T |_{\bm{x}=\bm{x}_m(t)} & = & \frac{\mathcal{L}}{c_p} \dot{\bm{x}}_m(t) 
\quad \forall \, \bm{x}_m(t) \in \mathcal{I}(t), \label{eq:bc2T}\\
\bm{u}|_{\bm{x}=\bm{x}_m(t)} & = & \bm{0}  \quad \forall \, \bm{x}_m(t) \in \mathcal{I}(t).
\end{eqnarray}
Here $\mathcal{L}$ and $c_p$ respectively are latent and specific heat; $\bm{x}_m(t)$ is the position vector 
of a point belonging to the interface (denoted $\mathcal{I}(t)$) and, consequently, $\dot{\bm{x}}_m(t)$ is 
the velocity at which the melting front advances into the solid.

The temperature equation (\ref{eq:T}) together with the associated phase-change boundary condition 
(\ref{eq:bc2T}) can be recast as follows:
\begin{equation}\label{eq:Heat4}
\partial_t T + \bm{u} \cdot \grad T = \kappa \bm{\nabla}^2 T-\frac{\mathcal{L}}{c_p}\ \partial_t \phi_l ,
\end{equation}
where the phase field $\phi_l(\bm{x},t)$ accounts for 
the volume fraction of the liquid phase ($\phi_l=0$ in the solid and $\phi_l=1$ in the fluid).
Such a formulation, can be derived from the transport equation for the enthalpy field $\mathcal{H}(\bm{x};t) =c_p T+\phi_l \mathcal{L}$,  which is the sum of the sensible heat and the latent heat associated to the phase-change process, \cite{shyy}.
Finally, we note that the local instantaneous height of the liquid melt layer 
can be obtained from $\phi_l$: 
\begin{equation}
z_m(x,y,t)=\int_0^{H_{max}} \phi_l(\bm{x},t)\ dz.
\label{eq:interfaceheight}
\end{equation}

\subsection{Control parameters \label{sec:parameters}}
It is convenient to express the equations of motion of the system in dimensionless form.
We define non-dimensional variables by dividing temperature by 
$\Delta T=T_{0}-T_{m}$, density by $\rho_0$, length by $H_{max}$ and time 
by the diffusive time $H_{max}^2/\kappa$. The evolution equations (\ref{eq:vel}) and (\ref{eq:T}), with condition (\ref{eq:nodivergence}),  
made non-dimensional read:
\begin{eqnarray}
\frac{\partial \tilde{\bm{u}}}{\partial \tilde{t}}+
\tilde{\bm{u}} \cdot \tilde{\bm{\nabla}}\tilde{\bm{u}} & = & 
-\tilde{\bm{\nabla}} \tilde{p} + Pr \tilde{\bm{\nabla}}^2\tilde{\bm{u}}+Ra_{max}\ Pr\ \tilde{T} \bm{\hat{\tilde{z}}}, 
\label{eq:vel_nondim}\\
\tilde{\bm{\nabla}} \cdot \tilde{\bm{u}} & = & 0, \label{eq:nodiv_nondim}\\
\frac{\partial \tilde{T}}{\partial \tilde{t}} + 
\tilde{\bm{u}} \cdot \tilde{\bm{\nabla}} \tilde{T} & = & \tilde{\bm{\nabla}}^2 \tilde{T} 
-\frac{1}{St}\frac{\partial \phi_l}{\partial \tilde{t}}, \label{eq:T_nondim}
\end{eqnarray}
with tildes indicating non-dimensional variables (note however that $\tilde{p}$ is a modified dimensionless pressure 
that, differently from $p$, does not include the hydrostatic component and whose gradient reabsorbs other constant terms).  
In Eqs.~(\ref{eq:vel_nondim}-\ref{eq:T_nondim}) three global control parameters appear: 
\begin{equation} 
Ra_{max} = \frac{\beta g \Delta T H_{max}^3}{\nu \kappa},  \qquad  Pr=\frac{\nu}{\kappa}, \qquad  St=\frac{c_{p}\Delta T}{\mathcal{L}}.
\end{equation}
They respectively are: the Rayleigh number ($Ra_{max}$), accounting for the relative strength of 
buoyant forcing and dissipative processes; the Prandtl number ($Pr$), 
expressing the ratio of kinematic viscosity ($\nu\equiv\mu/\rho_0$) to thermal diffusivity; the Stefan number ($St$), 
giving the ratio of sensible to latent heat. 
Note that with the latter definition, the singular limit $St \to 0$ represents the case of a material that needs an infinite time to melt.

In the present study we are interested in the dynamics of the system before the melting interface reaches the top boundary. 
This, combined with the fact that the solid is initially at the melting temperature, means that $H_{max}$ is not 
a characteristic scale of the problem. In fact, it plays no role here, given that there is 
neither thermal advection nor
diffusion in the solid phase.
For this reason it is more convenient to adopt as a reference length scale the instantaneous 
horizontally averaged height of the fluid layer, $H(t)$, which is defined as
\begin{equation}
H(t) =  \frac{1}{L^2} \int_0^L \int_0^L z_m(x,y,t) dx \ dy 
\end{equation}
or, equivalently,
\begin{equation}
H(t) = \frac{1}{L^2} \int_V \phi_l\ d^3x  = H_{max} \langle \phi_l \rangle.
\label{eq:H_phi}
\end{equation}
In Eq.~(\ref{eq:H_phi}), the notation $\langle \ldots \rangle \equiv V^{-1} \int_V \ldots d^3 x$ indicates 
a volume average over the entire domain (i.e. fluid and solid); 
hence $\langle \phi_l \rangle$ denotes the global liquid fraction in the system.
This allows to introduce the effective Rayleigh number:
\begin{equation}
Ra_{eff}=\frac{\beta g \Delta T [H(t)]^3}{\nu \kappa} = Ra_{max}\ \langle \phi_l \rangle^3.
\label{eq:raeff}
\end{equation}
A further control parameter characterizing the system is the geometrical aspect ratio. Also in this case it makes sense to define an effective aspect ratio,
\begin{equation}
\Gamma_{eff} = \frac{L}{H(t)} = \frac{L}{H_{max} \ \langle \phi_l \rangle}   =  \frac{\Gamma_{min}}{ \langle \phi_l \rangle} 
\label{gammaeff}
\end{equation}
Note that during its dynamics the convective melting system always explores a range of decreasing effective aspect ratios, 
starting from $\Gamma_{eff} = + \infty$  and reaching a value than cannot be smaller than $\Gamma_{min}  \equiv L / H_{max}$.

\section{Heat flux \label{sec:math}}
In this section we derive the global relations expressing the vertical heat flux across the fluid layer. 
We shall distinguish between the heat flux at the bottom side of the system, that we will call incoming flux, 
and the heat flux at the fluid-solid interface, that we will call outgoing flux.
Here we focus on the 3D configuration. However, the developed arguments can be adapted to the 2D case with no 
conceptual difficulty.

\subsection{Global heat budget \label{sec:heatbudget}}
We begin by considering the equation for temperature in the fluid domain with moving interface, Eq. (\ref{eq:T}), in 3D. 
Writing it in conservative form and integrating over the volume $V_l$ occupied by the fluid, one obtains 
\begin{equation}
\int_{V_l}\partial_t T d^3x + \int_{\partial V_l} \bm{n}\cdot(\bm{u} T -\kappa\grad{T}) dS=0,
\end{equation}
after making use of the divergence theorem and where $\bm{n}$ denotes the outward pointing unit normal vector associated 
with the orientation of the surface $\partial V_l$.
Due to the velocity and temperature boundary conditions (Sec. \ref{sec:eqs_motion}), the contribution from 
the advective term $\bm{n} \cdot \bm{u} T$ is zero and one is left with: 
\begin{gather}
\int_{V_l}\partial_t T d^3x + \int_0^L \int_0^L \kappa\partial_z T|_{z=0}dx~dy +
\int_{\mathcal{I}} -\kappa\bm{n}\cdot\grad{T}|_{\bm{x}=\bm{x}_m(t)} dS=0,
\end{gather}
where the first surface integral is evaluated at the bottom flat boundary and the second at the melting front. 
After normalizing each term by the horizontal bottom surface ($L^2$) and rearranging, one gets:
\begin{eqnarray}
Q^{in} & = & Q^{out}+L^{-2}\int_{V_l}\partial_t T d^3x, \label{eq:Heat_Flux} \\
Q^{in} & = & -\kappa\left<\partial_z T  |_{z=0} \right>_{A}, \label{eq:Qin} \\
Q^{out} & = & L^{-2}\int_{\mathcal{I}} -\kappa\bm{n}\cdot\grad{T}|_{\bm{x}=\bm{x}_m(t)} dS, \label{eq:Qout}
\end{eqnarray}
where $\langle \ldots \rangle_A$ stands for an average over a horizontal plane. In the above expressions, $Q^{in}$ can be identified 
with the bottom heat flux (incoming into the fluid), expressed in $\rho_0 c_p$ units, and $Q^{out}$ with the heat flux at the top of the fluid domain (outgoing into 
the solid), in the same units. The last term in Eq. (\ref{eq:Heat_Flux}) expresses the total temporal variation of the temperature in the melt and it therefore represents the global heating 
of the system. It results from the non-stationarity of the dynamics; in the RB system this term vanishes when a time average is also 
performed.
Equation (\ref{eq:Heat_Flux}) can be recast in terms of the dimensionless Nusselt number, normalizing by $\kappa\Delta T/H_{max}$. 
This gives
\begin{equation}
Nu^{in}=Nu^{out}+\frac{H_{max}}{\kappa\Delta T}\frac{1}{L^2}\int_{V_l}\partial_t T d^3x.
\label{eq:Nusselt}
\end{equation}
As in Sec. \ref{sec:parameters}, it seems here convenient to introduce an effective Nusselt number:
 \begin{equation}
 Nu_{eff} =  Nu\ \frac{H(t)}{H_{max}} = Nu\ \langle \phi_l \rangle.
 \label{eq:Nusselt_eff}
\end{equation}
The meaning of the effective Nusselt number is the usual one. It expresses the ratio between the total heat flux and the 
one that would take place across the scale $H(t)$ with a temperature gap $\Delta T$ in a stationary process controlled by 
diffusivity ($\kappa$) only. We note that this way of normalizing the heat flux was previously introduced in \cite{Ulvrova}.
The non-dimensional version of Eq. (\ref{eq:Nusselt}) is
\begin{equation}
Nu^{in}_{eff}=Nu^{out}_{eff}+ \left<\phi_l\right>^2 \langle \partial_{\tilde{t}} \tilde{T} \rangle_{V_l},
\label{eq:Nu_eff_nondim}
\end{equation} 
where $\langle \ldots \rangle_{V_l}$ indicates an average over the liquid volume $V_l=H(t)L^2$.

To better appreciate the meaning of the term $Nu^{out}_{eff}$ in our system, let us consider the 
temperature equation, Eq. (\ref{eq:Heat4}) in the full (liquid and solid) domain:
\begin{equation}
\partial_t T + \nabla \cdot ( \textbf{u}  T - \kappa\ \nabla T ) = - \frac{\mathcal{L}}{c_p} \partial_t \phi_l .
\label{eq:T_cons_phi}
\end{equation}
Proceeding as before, but now with volume integrals over the whole domain, we obtain
\begin{equation}
Nu_{eff}^{in}=\frac{1}{St} \frac{H_{max}^2}{\kappa} \left<\phi_l\right> \frac{d\left<\phi_l\right>}{dt}
+\frac{H(t)}{\kappa\Delta T}\frac{1}{L^2}\int_{V}\partial_t T d^3x
\label{eq:Nu_eff_in}
\end{equation}
and in dimensionless units:
\begin{equation}
Nu_{eff}^{in}=\frac{1}{St}  \left<\phi_l\right> \frac{d\left<\phi_l\right>}{d\tilde{t}} + 
\langle \phi_l \rangle \langle \partial_{\tilde{t}} \tilde{T} \rangle.
\label{eq:Nu_eff_in_nondim}
\end{equation}
Observing, now, that $ \langle \ldots \rangle = \langle \phi_l \rangle \langle \ldots \rangle_{V_l} + 
(1-\langle \phi_l \rangle) \langle \ldots \rangle_{V_s}$, where $\langle \ldots \rangle_{V_s}$ is the average 
over the volume of the solid phase, and using this to transform the last term in Eq. (\ref{eq:Nu_eff_in_nondim}), we get: 
\begin{equation}
Nu_{eff}^{in}=\frac{1}{2 St}  \frac{d\lra{\phi_l}^2}{d\tilde{t}}+ \lra{\phi_l}^2   \lra{\partial_{\tilde{t}} \tilde{T}}_{V_l} 
            + (\lra{\phi_l}  - \lra{\phi_l}^2) \lra{\partial_{\tilde{t}} \tilde{T}}_{V_s}.
\label{eq:Nu_eff_in_phi}
\end{equation}
In the special case in which the solid is initially uniformly at the melting temperature $T_m$, no conduction occurs in the 
solid phase and hence $\lra{\partial_{\tilde{t}} \tilde{T}}_{V_s} = 0$. 
Comparing the above equation with Eq. (\ref{eq:Nu_eff_nondim}) one then recognizes that the outgoing heat flux is directly linked 
to the first term on the right-hand side of Eq. (\ref{eq:Nu_eff_in_phi}), proportional to the melt fraction variation over time.

In summary, when the solid is initially at the melting temperature we have:
\begin{eqnarray}
Nu_{eff}^{in} & = & -\left<\partial_{\tilde{z}} \tilde{T}  |_{\tilde{z}=0} \right>_{A}  \left<\phi_l\right>,  \label{eq:summarynu_in}\\
Nu_{eff}^{out} & = & \frac{1}{2 St} \frac{d\left<\phi_l\right>^2}{d\tilde{t}}, \label{eq:summarynu_out}\\
Nu_{eff}^{in} - Nu_{eff}^{out} & = & \left<\phi_l\right>^2  \langle \partial_{\tilde{t}} \tilde{T}\rangle_{V_l} > 0. \label{eq:Nuoutnuin}
\end{eqnarray}
Equation (\ref{eq:summarynu_in}) is the non-dimensional analogue of Eq. (\ref{eq:Qin}) for the incoming heat flux, and 
Eq. (\ref{eq:summarynu_out}) relates the outgoing heat flux to the liquid fraction, and in particular to the global melting rate, defined as $d \langle \phi_l \rangle / d\tilde{t}$.
The last inequality, in Eq. (\ref{eq:Nuoutnuin}), follows from the fact that one expects that not all the heat input into the system will 
be transferred across the fluid layer, but that part of it will be used to warm the liquid to an intermediate temperature 
between the minimum value $T_m$ and the maximum one $T_0$.

\subsection{Scaling relations for the heat flux and the melt fraction \label{sec:scaling}}
Before the onset of convection, i.e. for time small enough, the system evolution is governed by heat conduction 
in the fluid layer and melting at its boundary with the solid. 
In such conditions, the liquid-solid interface is flat; the incoming and outgoing heat fluxes are respectively given by   
\begin{equation}
Nu_{eff}^{in} = \frac{2 \lambda^2}{St}\ e^{\lambda^2} \qquad\textrm{and} \qquad  Nu_{eff}^{out} = \frac{2 \lambda^2}{St},
\label{eq:heatflux_conduction}
\end{equation}
where $\lambda$ is a constant depending on $St$ \cite{alexiades1992mathematical}. Further details can be found in appendix \ref{sec:conductive}.  
Here we only note that both $Nu_{eff}^{in}$ and $Nu_{eff}^{out}$ are time independent. 
In the limit of small $St$ it is possible to show that $\lambda \simeq \sqrt{St/2}$ and, 
therefore, $Nu_{eff}^{in} \simeq 1 + St/2$ and $Nu_{eff}^{out} \simeq 1$.

In the convective regime, due to the important non-linearities of the dynamics, the exact expression  
of the liquid fraction as a function of time, and hence of the heat fluxes, is not available. However, 
Eqs. (\ref{eq:summarynu_in}-\ref{eq:Nuoutnuin}) can still be used to extract informative scaling relations. 
Similarly to what is done for turbulent convection in the RB system, one can assume that the effective Nusselt number 
has power-law dependencies on the control parameters of the system. 
We will here consider the outgoing effective Nusselt number and assume: 
\begin{equation}
Nu_{eff}^{out} \sim Ra_{eff}^{\alpha}\ Pr^{\delta}\ St^{\gamma}.
\label{eq:Nu_eff_scaling}
\end{equation}
By using Eq. (\ref{eq:summarynu_out}) to relate $Nu_{eff}^{out}$ to $\langle \phi_l \rangle$, as well as 
the definition of $Ra_{eff}$ (Eq. (\ref{eq:raeff})), we then obtain the following scaling for the melt fraction:
\begin{equation}
\langle \phi_l \rangle \sim \tilde{t}^{\frac{1}{2-3\alpha}}\ Pr^{\frac{\delta}{2-3\alpha}}\ St^{\frac{\gamma+ 1}{2-3\alpha}}.
\label{eq:phi-scaling}
\end{equation}

Few observations are in order. First, in the conductive case, because $Nu_{eff}^{out}$ is constant, 
one has $\alpha= \delta = \gamma = 0$ and  Eq. (\ref{eq:phi-scaling}) gives the known behavior 
(see appendix \ref{sec:conductive}) $\langle \phi_l \rangle \sim \tilde{t}^{1/2}\ St^{1/2}$ (where the 
limit of small $St$ has been taken too). Second, in the presence of convection, again by analogy with the RB system, 
the value $\alpha=1/3$ may be considered. We remind that in this context the $1/3$ Rayleigh exponent corresponds to the 
so-called Malkus scaling \cite{malkus_1954}, a regime where the horizontal thermal boundary layers are marginally stable or 
in which the vertical heat flux does not depend on height \cite{Chilla}. In the CM context the same scaling exponent, 
this time for the effective Rayleigh, corresponds to a time independent average melting front speed 
$ v_m \propto \frac{d}{d\tilde{t}} \left<\phi_l\right> = \mathrm{const}$.
Third, the so-called ultimate regime of thermal convection, which is dominated by the flow dynamics in the bulk of the system 
and is characterized by $\alpha=1/2$  and $\delta=1/2$, would give: $ v_m \sim Pr\ \tilde{t}$, i.e. constant front acceleration. 
Finally we note that little is known about the Stefan dependency of the global heat flux in the convective regime. However, 
we can observe that if $Nu_{eff}^{out}$ is independent of $St$ ($\gamma=0$) and at the same time $\alpha=1/3$, 
$v_m$ would linearly depend on $St$, $ v_m \sim St$.

\section{Methods \label{sec:methods}}
We perform direct numerical simulations (DNS) of the convective melting system introduced in Sec. \ref{sec:model} and, 
for comparison purposes, of thermal convection between fixed flat parallel plates, i.e. of the RB system. 
Different methods have been proposed in the past to simulate melting coupled to flows. They can be grouped into two main 
classes: front-tracking (moving boundary) methods and single-domain fixed-grid ones (as, e.g., the enthalpy method).  
Both types of approaches clearly have advantages and drawbacks; if front-tracking generally allows for a smoother 
resolution of the interface, the enthalpy method typically lends to simpler implementations, particularly in 3D setups.
Our simulations are based on a uniform mesh Lattice-Boltzmann (LB) method \cite{succi2001lattice} employing an enthalpy formulation, 
similar to the one proposed in \cite{Huber}, to discretize Eq. (\ref{eq:Heat4}). The technical details of the numerics are discussed 
in appendix~\ref{sec:numerics}.

Having in mind melt ponds in the Arctic, we focus on water-ice dynamics. For this reason we keep the Prandtl number fixed at 
the value $Pr=10$, close to that of fresh water just above the freezing point (at temperatures $0.01^\circ C<T<10^\circ C$, 
Prandtl is $9.47<Pr<13.67$) \cite{Skyllingstad}. 
Ulvrova and colleagues \cite{Ulvrova} addressed the same melting problem but for $Pr=7$, $St=0.9$ and for $Pr = \infty$, $St=10$; 
the former case can be representative of water at $20^\circ C$ while the latter seems a reasonable approximation for 
convection in the solidifying Earth's mantle. 
The $O(1)$ value of the Stefan number used in the study mentioned above is advantageous for numerical computations 
but not always realistic for geophysical applications. For instance, in ice melt ponds $St$ is estimated to be 
$O(10^{-2})$.
For computational reasons, in the present work we also take $St=1$ in the majority of simulations, 
but we will also present results of computationally more expensive simulations at $St=10^{-1}$ or cheaper (faster) 
ones at $St=10$ and $100$.

The simulations are initialized by setting the fluid fraction to zero in the whole domain and temperature at the 
melting value $T=T_m$. A small random perturbation (of amplitude $T_{\epsilon} \leq 10^{-6}$) is added to destabilize the system,  
which is known to be linearly unstable \cite{Kim2008}, as in the RB case. 
Ensemble averages are performed over several simulations with different random initial conditions.  
In table \ref{tab:table1} we summarize the most relevant information on all the CM simulations performed,  
listing both the numerical parameters adopted in the LB simulations and the resulting dimensionless control parameters. 
To guide the reader we also provide an indication on where the obtained data are employed in the figures of the paper.

\begin{table}[t]
\begin{tabular}{|c|c|c c c| c c c c c c c|c c c c|c|}
\hline
 &\begin{tabular}{@{}c@{}} \textit{n. of runs}\end{tabular}&$L_x$&$L_z$&$L_y$&$\nu$&$\kappa$&$\beta$&$\Delta T$ & $g$ &\begin{tabular}{@{}c@{}} $\mathcal{L}$ \end{tabular}&\begin{tabular}{@{}c@{}} $c_p$\end{tabular}&$Pr$&$St$&$Ra_{max}$& $\Gamma_{min}$ &\textit{Fig. n.}\\
\hline
\multirow{5}{*}{2D}
&8 & 2000 & 1000 & 1 & 0.2 & 0.02 & 0.0005 & 1 & 1 & 1 & 100 & 10 & 100 &  $1.25\times10^{8}$& 2 & \ref{fig:fluxinout} \\
\cline{2-17}
&8 & 2000 & 1000 & 1 & 0.2 & 0.02 & 0.0005 & 1 & 1 & 1 & 10 & 10 & 10 &  $1.25\times10^{8}$& 2 & 
\ref{fig:stdependency1}, \ref{fig:fluxinout} \\
\cline{2-17}
&8 & 2000 & 1000 & 1 & 0.2 & 0.02 & 0.0005 & 1 & 1 & 1 & 1 & 10 & 1 &  $1.25\times10^{8}$& 2 &
\begin{tabular}{c}
\ref{fig:visualization2d},  \ref{fig:aspect}, \ref{fig:aspect3}, \ref{fig:comparison}, \ref{fig:reynolds},\\ \ref{fig:cor_dev}, \ref{fig:stdependency1}, \ref{fig:fluxinout}
\end{tabular}\\
\cline{2-17}
&8 & 2000 & 1000 & 1 & 0.2 & 0.02 & 0.0005 & 1 & 1 & 10 & 1 & 10 & 0.1 &  $1.25\times10^{8}$& 2 &
\ref{fig:stdependency1}, \ref{fig:fluxinout} \\
\hline
\multirow{1}{*}{3D}
&6 & 512 & 512 & 512 & 0.2 & 0.02 & 0.003 & 1 & 1 & 1 & 1 & 10 & 1 &   $1.00\times10^{8}$& 1 &  \ref{fig:comparison}, \ref{fig:reynolds}, \ref{fig:visualization3d}, \ref{fig:cor_dev}, \ref{fig:stdependency1}, \ref{fig:fluxinout}\\
\hline
\end{tabular}
\caption{Summary of the parameter values for all CM simulations. We provide both dimensional (in simulation units) 
and dimensionless control parameters. The time step is fixed to $\delta t =1$ for all simulations. 
The mesh is uniform and the numerical value of the grid spacing is $\delta x = 1$; 
$L_x, L_y, L_z$ indicate the size of the system in mesh units.
The second column from the left specifies the number of replica simulations performed, which are employed to 
compute ensemble averages. 
The last column on the right indicates the figures of the paper where the results of the 
simulations are used.}
\label{tab:table1}
\end{table}

The RB simulations are performed with parameter values as close as possible to those of the CM case. The Rayleigh number is 
here set by controlling the height of the system ($L_z$) and we make sure to always have at least 8 grid points in the 
thermal boundary layer. Simulations ran over tens or hundreds of large-eddy turnover times. 
The turnover time is defined as $T_e = L/\overline{u_{rms}}$ with $L$ the width of the system, 
$u_{rms}=\sqrt{ \langle |\bm{u}|^2 \rangle_{V}}$, and the overbar denoting a temporal average 

Note that in the presence of melting the root-mean-square velocity $u_{rms}$ is computed differently. 
Indeed, due to the unsteady character of the flow it would be now inappropriate to average over time. Moreover, because there is no 
velocity in the undeformable solid, it makes more sense to compute the spatial average over the fluid domain only. 
Accordingly, one has:
\begin{equation}
u_{rms,V_l}=\sqrt{ \langle |\bm{u}|^2} \rangle_{V_l}  = \sqrt{ \langle |\bm{u}|^2} \rangle_{V} \ \langle \phi_l \rangle^{-1/2} = u_{rms} \ \langle \phi_l \rangle^{-1/2}.
\label{eq:urmsVl}
\end{equation}
This quantity will be used to construct the Reynolds number, which will be discussed in Sec. \ref{sec:results}.

\section{Results \label{sec:results}}

\subsection{Qualitative description of the dynamics \label{sec:qualitative}}
\begin{figure}[!htb]
\begin{center}
\subfigure[$Ra_{eff}=5 \times 10^4$]{\includegraphics[width=1.\columnwidth]{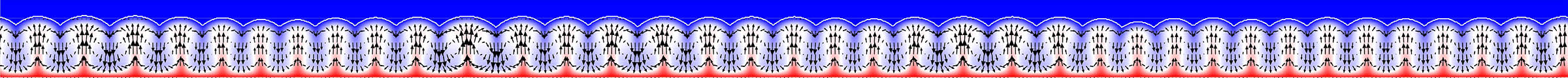}\label{fig:visualization2d-a}}
\subfigure[$Ra_{eff}=2.5\times 10^5$]{\includegraphics[width=1.\columnwidth]{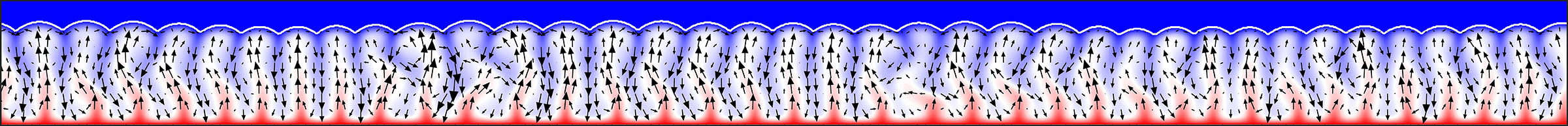}\label{fig:visualization2d-b}}
\subfigure[$Ra_{eff}=2.9\times 10^5$]{\includegraphics[width=1.\columnwidth]{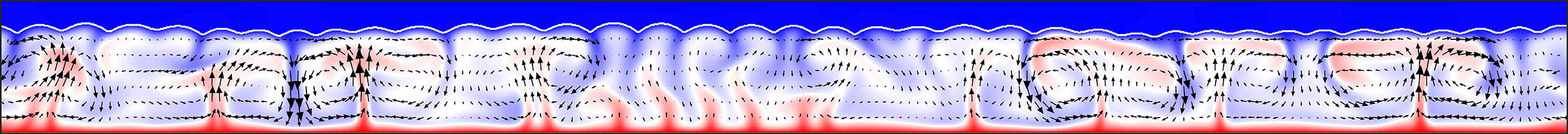}\label{fig:visualization2d-c}}
\subfigure[$Ra_{eff}= 5\times 10^5$]{\includegraphics[width=1.\columnwidth]{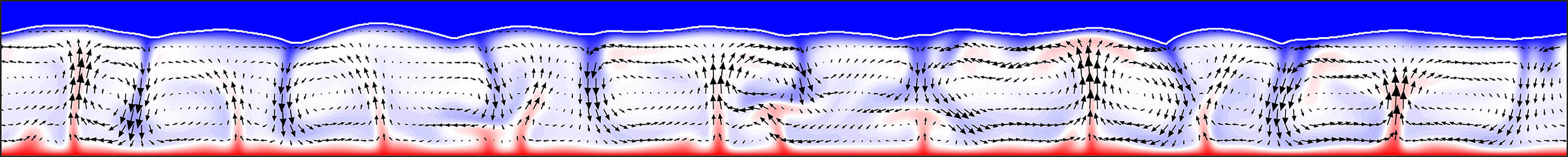}\label{fig:visualization2d-e}}
\subfigure[$Ra_{eff}=5\times 10^6$]{\includegraphics[width=1.\columnwidth]{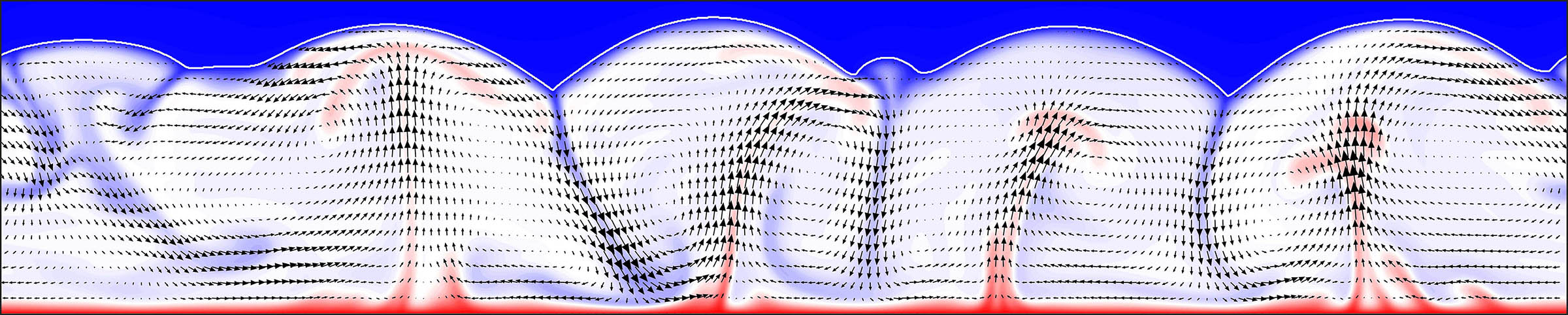}\label{fig:visualization2d-f}}
\subfigure[$Ra_{eff}=5\times 10^7$]{\includegraphics[width=1.\columnwidth]{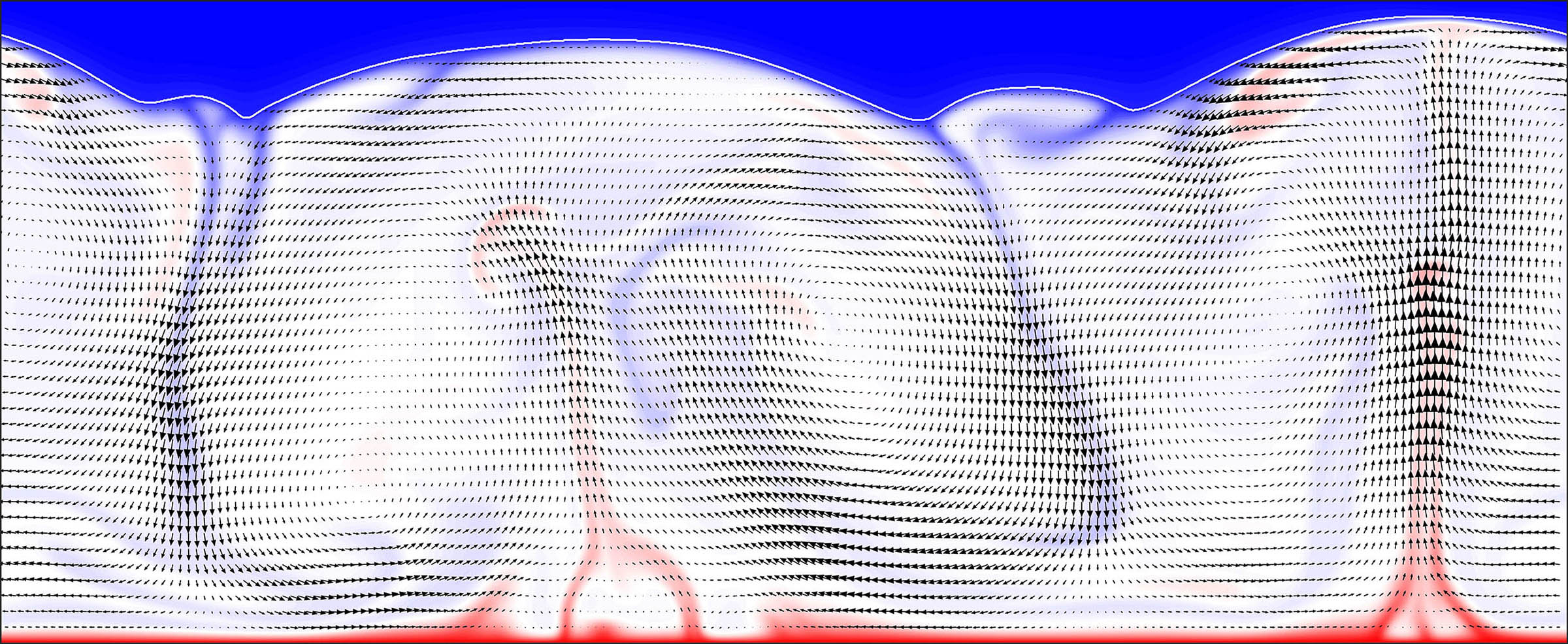}\label{fig:visualization2d-g}}
\caption{Visualization of the 2D CM system in different stages at increasing effective Rayleigh numbers (from top to bottom). 
The lateral size of the system is $2 H_{max}$ (or $L_x =2000$ in numerical units), while the vertical size has been cut at 
$\sim 1.1\ H(t)$. 
Color codes the normalized temperature deviation from the mean $(T_0+T_m)/2$ (red is for $-0.5$, blue for $0.5$ and white for zero). 
Arrows represent the velocity field normalized by the instantaneous maximum velocity magnitude. The thin white line 
in the top part of each temperature field marks the liquid-solid interface. The global parameters are $St=1$ and $Pr=10$.
A movie is available in the SM \cite{SM}.}
\label{fig:visualization2d}
\end{center}
\end{figure}

We start by describing the typical evolution of the CM model system. This passes through different stages. 
At the beginning, the melt layer grows solely by conduction and the system closely follows the Stefan solution.
There is no noticeable fluid flow and the phase-change interface remains flat. At later times a convective flow pattern develops.  
The onset of convection is delayed for increasing values of $St$, as in \cite{Kim2008}; the critical effective Rayleigh number 
recovers that of the RB system only in the limit of vanishing $St$ \cite{vasil_proctor_2011}. The convective onset occurs at around 
$Ra_{eff} \simeq 5\times 10^3$ in our 2D simulations. 
The flow visualization of Fig. \ref{fig:visualization2d} helps in elucidating the main features of the CM dynamics, 
from the onset stage on, in the 2D setting. 
The appearance of convection is marked by a change in the shape of the phase-change interface, from flat to a nearly periodic 
curve that is associated with the formation of transversal rolls (Fig. \ref{fig:visualization2d}a at $Ra_{eff}=5\times 10^4$).  
In this phase the convective rolls grow as if they were vertically stretched. This stage resembles the steady convection 
observed immediately after the onset in the RB system, except for the vertical growth of the fluid layer. 
At $Ra_{eff} = 2.5 \times 10^5$ (Fig. \ref{fig:visualization2d}b), the rolls start to oscillate laterally and, 
when the oscillations are large enough, they can merge in pairs. 
This has a repercussion on the interfacial curve, which is subsequently shaped by the new flow pattern after a 
time delay. Indeed at $Ra_{eff} = 2.9 \times 10^5$ (Fig. \ref{fig:visualization2d}c), we see that all the rolls have merged, 
creating two or three times wider convection cells, while the interface shape is not strongly affected yet. 
At $Ra_{eff} = 5 \times 10^5$ (Fig. \ref{fig:visualization2d}d), the interface loses periodicity and becomes smoother 
(i.e. without cusps). At $Ra_{eff} = 5 \times 10^6$ (Fig. \ref{fig:visualization2d}e) larger convective flow patterns 
establish. They are approximately twice as wide as those occurring at ten times smaller $Ra$ values. 
The interface now has cusps again and it is evident that such special points pin the detachment of cold plumes. 
For even larger $Ra_{eff}$ the convective cells are bigger and strongly fluctuate in time and space 
(Fig. \ref{fig:visualization2d}f at $ Ra_{eff} = 5 \times 10^7$). 

\subsection{Scaling in 2D \label{sec:effect2d}}
In order to address the quantitative features of the dynamics we study the intensity of the 
heat flux $Nu_{eff}^{in}$ as a function of the imposed forcing, here parameterized by $Ra_{eff}$.
The rationale for the choice of the incoming heat flux instead of the outgoing one is that it facilitates the comparison with 
the RB system, where the heat flux can be computed exactly the same way, Eq. (\ref{eq:summarynu_in}). Furthermore, for numerical 
reasons, the computation of $Nu_{eff}^{in}$ in the CM simulations is less affected by numerical noise. We will come back to the 
discussion of the differences between $Nu_{eff}^{in}$ and $Nu_{eff}^{out}$ at the end of Sec. \ref{sec:stdependency}. 
 
\begin{figure}[!htb]
\centering
\includegraphics[width=0.7\columnwidth]{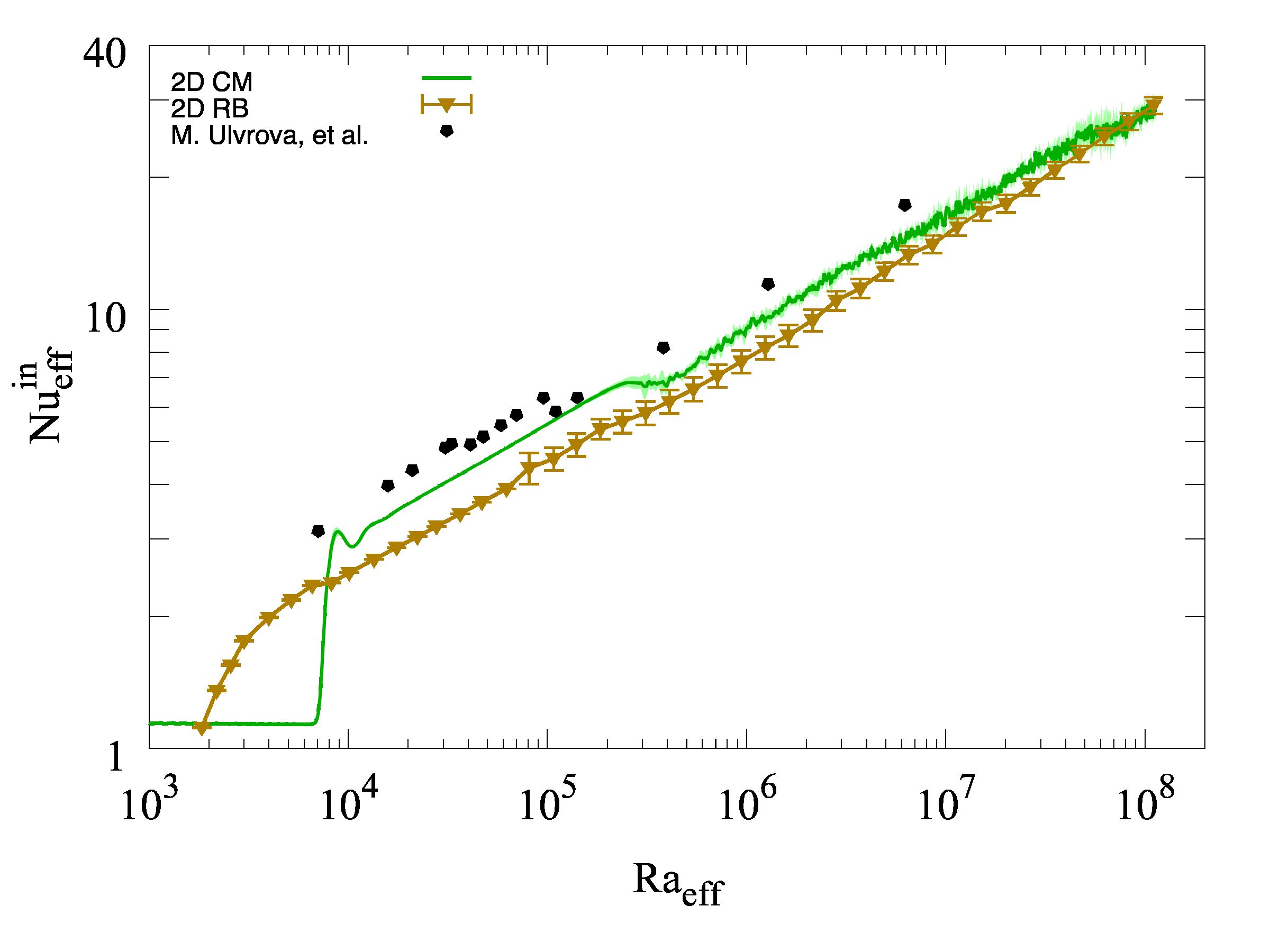}
\caption{Incoming dimensionless heat flux $Nu_{eff}^{in}$ as a function of the effective Rayleigh number $Ra_{eff}$  
for the CM and RB 2D systems. The minimum aspect ratio is $\Gamma_{min} \equiv L/H_{max} = 2$, meaning that the 
domain's width is always much larger than its height. The other CM control parameters are $St=1$ and $Pr=10$. 
In the RB case, $Pr=10$, too; the aspect ratio is chosen to always match the corresponding one, $\Gamma_{eff}$, of the CM system. 
Numerical results by Ulvrova and collaborators \cite{Ulvrova} are also shown; in this case the domain is laterally bounded, 
$Pr=7$ and $St=1$.}
\label{fig:aspect}
\end{figure}

Figure \ref{fig:aspect} shows $Nu_{eff}^{in}$ for both the 2D CM and the 2D RB systems. 
We can observe that the onset of convection in the CM system happens at higher Rayleigh number ($Ra_{eff} \simeq 5 \times 10^3$) than in the RB one ($Ra \simeq 1708$). Furthermore, a sudden jump in $Nu_{eff}^{in}$ is observed when convection is triggered in the CM system. 
A weakly nonlinear stability analysis based on a vanishingly small Stefan number assumption has been proposed for the CM system in \cite{vasil_proctor_2011}. In this work, a set of non-autonomous envelope equations was derived to describe the evolution of perturbations, which allowed to predict that the system bifurcates with a super-exponential amplitude growth at the onset of convection, occurring at $Ra_{eff} \simeq Ra_c$. Such a rapid growth is followed by a slower algebraic one, resulting from a saturation mechanism in a weakly nonlinear regime.  Although we cannot make a direct comparison with the results in  \cite{vasil_proctor_2011}, valid in the limit of vanishing Stefan number, such predictions are qualitatively consistent with the jump of $Nu_{eff}^{in}$ detected in our simulations at finite $St$ number.
Apart from the convective onset, and a relatively small amplitude mismatch ($Nu$ for CM is larger by at most $20\%$ with respect to $RB$)  the trends are very similar and the actual values of $Nu$ tend to be indistinguishable as $Ra_{eff}$ is increased.
The same figure reports the numerical results of \cite{Ulvrova} that, despite the different conditions (value of $Pr$, 
non-periodic lateral boundary conditions, initial temperature of the solid lower than the melting one $T_m$), are also close 
to ours. As for the RB system, this similarity of results in different conditions attests the robustness of the $Nu$-$Ra$ 
relation also in the CM system. 

\begin{figure}[!htb]
\begin{center}
\includegraphics[width=0.7\columnwidth]{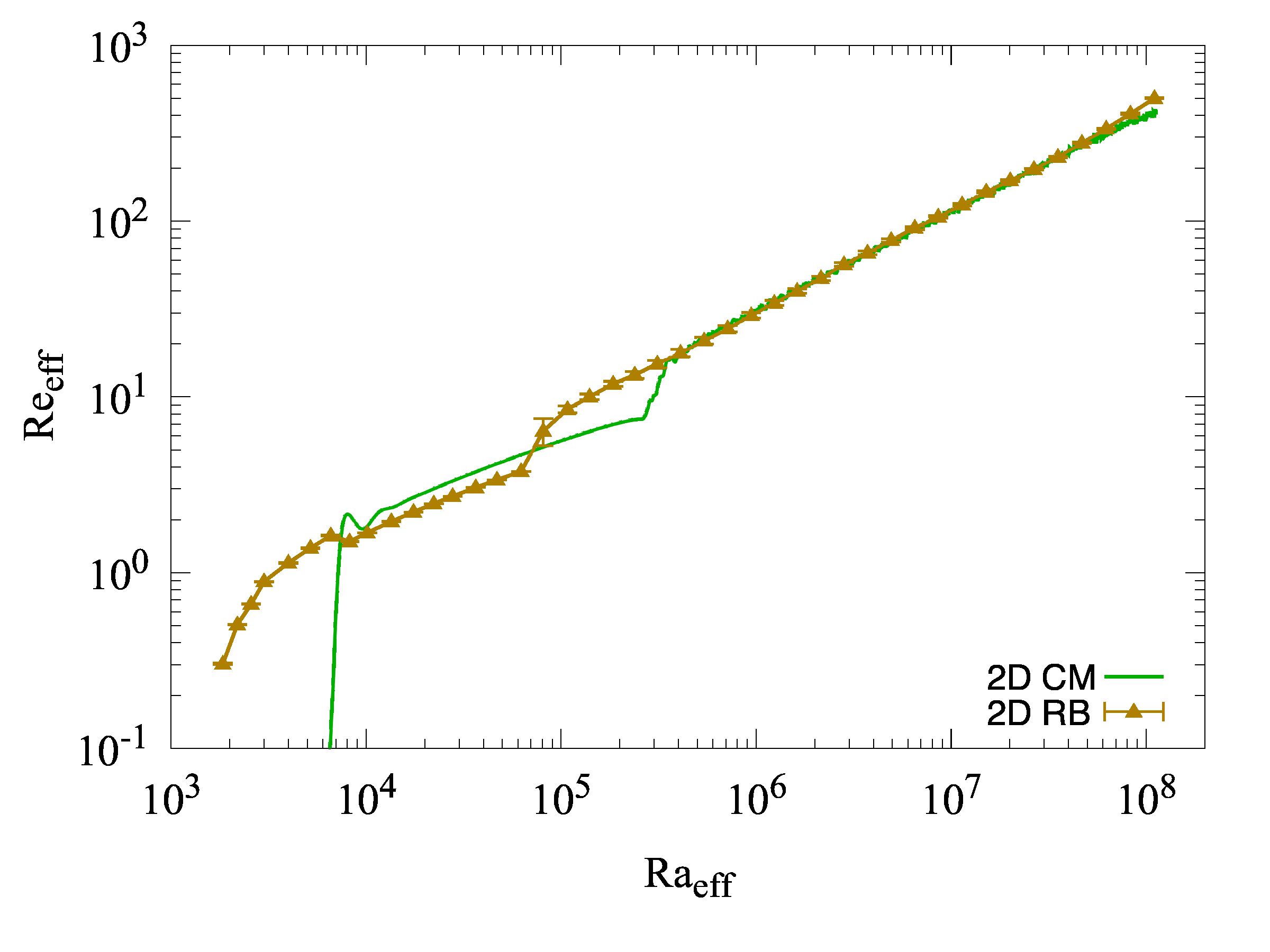}
\end{center}
\caption{
Effective Reynolds number versus the effective Rayleigh number for the CM and RB 2D systems. In both cases the control 
parameters are $St=1$ and $Pr=10$.}
\label{fig:aspect3}
\end{figure}

To complement this picture we also look at the scaling of kinetic energy. Reasoning in terms of dimensionless variables, 
this amounts to consider an effective Reynolds number:
\begin{equation}
Re_{eff}= \frac{u_{rms,V_l}\ H(t)}{\nu} = \frac{u_{rms} \langle \phi_l \rangle^{1/2} H_{max} }{\nu}, 
\label{eq:reynolds}
\end{equation}
where $u_{rms,V_l}$ is the fluid root-mean-square velocity computed over the liquid phase only 
(Eq. (\ref{eq:urmsVl})) and $u_{rms}$ its counterpart obtained from averaging over the whole volume. 
We observe here, Fig. \ref{fig:aspect3}, that the agreement between the CM and RB behaviors is remarkable, particularly in the 
range $Ra_{eff} \geq 4\times 10^{5}$. At lower $Ra_{eff}$ the differences can be ascribed to the delayed (in $Ra_{eff}$) 
transitions occurring when melting is present, as compared to the RB system. For instance, the transition from steady to 
laterally oscillating patterns occurs at around $Ra \simeq 6\times 10^4$ and $Ra_{eff} \simeq 3 \times 10^5$ in the RB and CM systems, 
respectively. 

A possible explanation for the fact that the magnitude and scaling of global quantities in the RB and CM systems are so close 
can be provided on the basis of a comparison between two characteristic velocity scales. The first one is the typical 
flow intensity $u_{rms,V_l}$, while the second one is the vertical melting front mean velocity $v_m = dH(t)/dt$. 
It is reasonable to conjecture that the CM system will behave as the RB one if the melting front moves slowly with respect to 
the flow: $v_m \ll u_{rms , V_l}$. This relation can be expressed in dimensionless form, via Eqs. (\ref{eq:summarynu_out}) 
and (\ref{eq:reynolds}):
 \begin{equation}
Nu_{eff}^{out} \ll \frac{Pr\ Re_{eff}}{St}. 
\label{eq:slowfront}
\end{equation}
According to Eq. (\ref{eq:Nuoutnuin}) we can expect $Nu_{eff}^{in}>Nu_{eff}^{out}$. Therefore, if the above condition 
is satisfied for the \textit{in} Nusselt number it will be satisfied also for the \textit{out} one. 
From Fig.~\ref{fig:aspect} we see that $2 \lesssim Nu_{eff}^{in} \lesssim 32$ for $10^4 \lesssim Ra_{eff} \lesssim 10^8$ and 
from Fig~\ref{fig:aspect3} that $1 \lesssim Re_{eff} \lesssim 500$ in the same range of Rayleigh numbers. Since here $Pr/St = 10$, 
it is then easily verified that indeed $Nu_{eff}^{in} \ll Re_{eff} Pr/St$ over this rather broad range of $Ra_{eff}$ values. Hence, 
$Nu_{eff}^{out}$ also fulfills condition (\ref{eq:slowfront}) essentially at all stages of the CM evolution, which indicates that the 
front speed is considerably smaller than the typical fluid velocity fluctuations and justifies the similarity with RB dynamics.

\subsection{Scaling in 3D \label{sec:effect3d}}
The differences in the functional behavior of global observables, such as the heat flux or the kinetic energy, between 2D and 3D 
RB convection have already been investigated in depth. 
Recently, the dynamics of laterally bounded 2D and 3D RB systems were compared in numerical simulations \cite{van2015comparison}. 
Exploring conditions corresponding to Rayleigh numbers up to $Ra=10^8$, with $0.045\le Pr \le 55$ and $\Gamma=1$, 
it was demonstrated that the 2D dimensionless global heat flux obeys the same scaling with $Ra$ as in 3D  
but for an approximately constant multiplicative factor, i.e., $Nu_{2D} \simeq K\cdot Nu_{3D}$ with $K<1$. 
The aim of this section is to assess to what extent this observation also holds for the CM system.

\begin{figure}[!htb]
\begin{center}
\subfigure[]{\includegraphics[width=0.7\columnwidth]{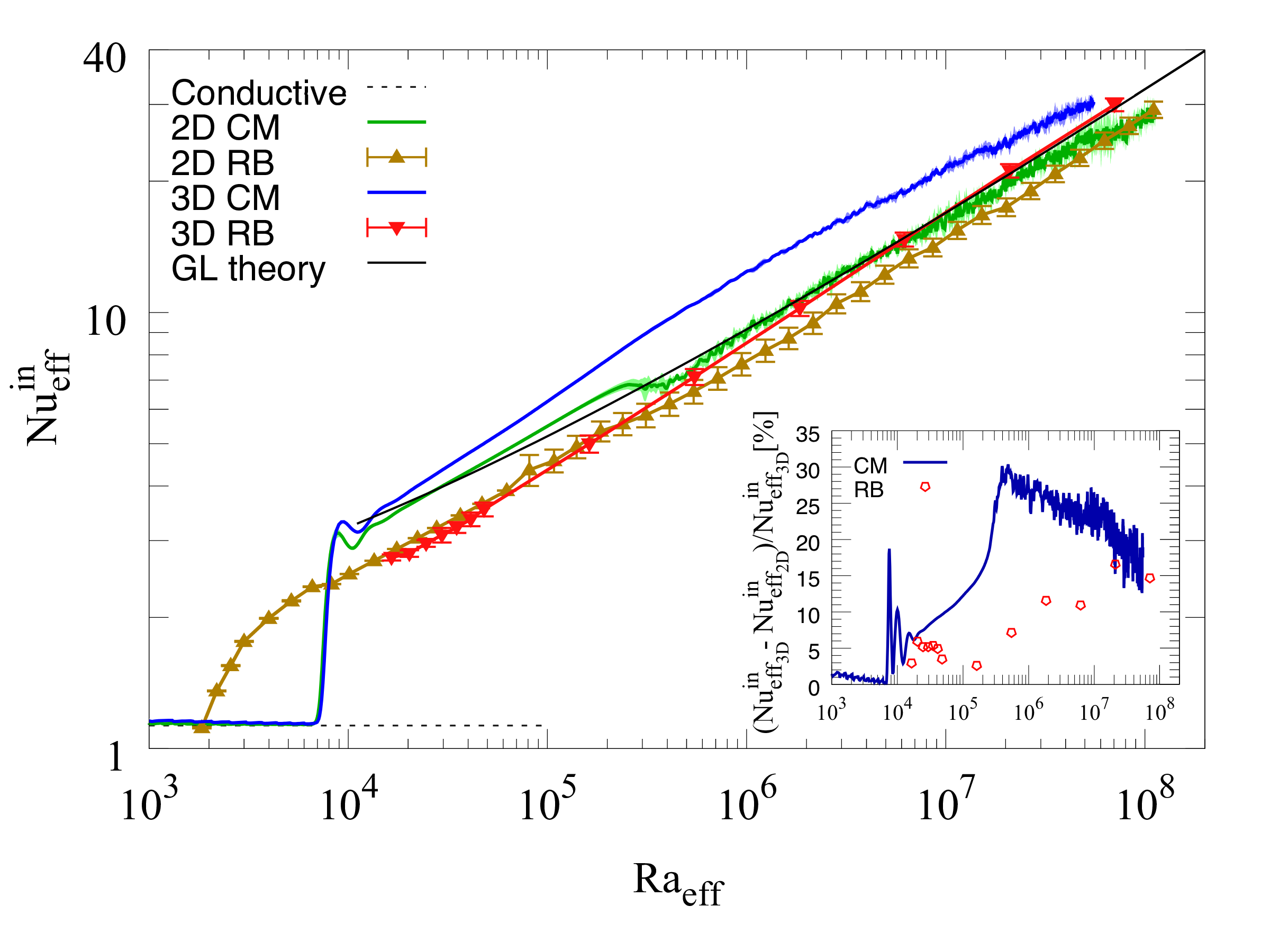}}
\subfigure[]{\includegraphics[width=0.7\columnwidth]{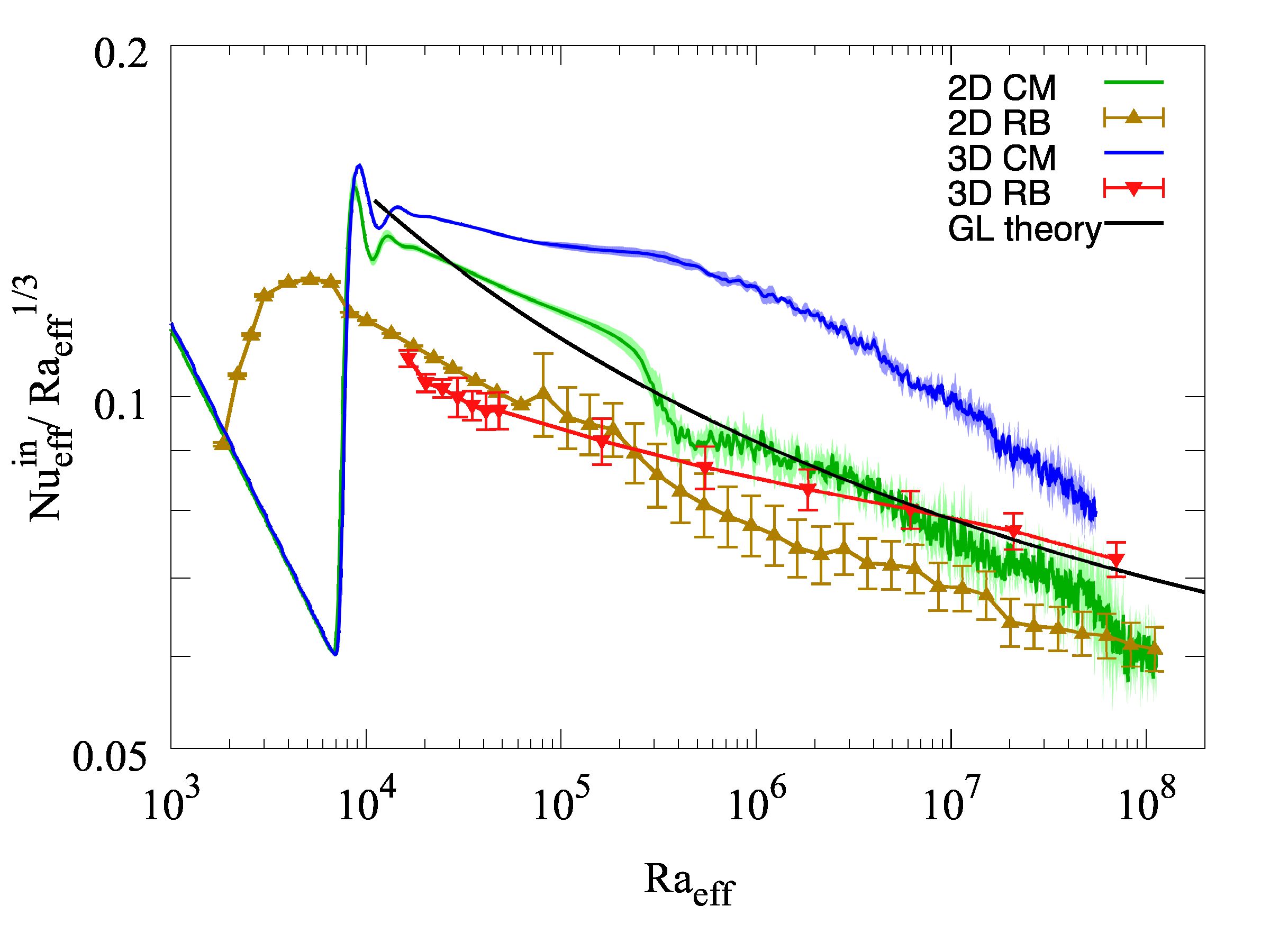}}
\caption{(a) Incoming effective Nusselt number versus the effective Rayleigh number 
for 2D and 3D systems with the same melting process' configuration; 
the global parameters are $St=1$ and $Pr=10$. 
The $Nu \, \mathrm{vs.}  \, Ra$ data for the RB 2D and 3D cases are also shown. The horizontal dashed line is the conductive value 
for the effective Nusselt number, Eq. (\ref{eq:heatflux_conduction}), while the solid black line is the prediction from 
GL theory, calculated as in \cite{grossmann_lohse_2000,stevens2013unifying}. The inset shows the relative difference (in percentage) 
between the 2D and 3D global heat fluxes for the CM and RB systems, for different Rayleigh numbers in the same range as in the main panel. 
(b) Same as (a) but for the Nusselt number compensated by $Ra_{eff}^{1/3}$.}
\label{fig:comparison}
\end{center}
\end{figure}

As a preliminary numerical test, we perform 2D and 3D RB simulations and check the $Nu-Ra$ relation for laterally periodic systems. 
We also check the agreement with Grossmann-Lohse (GL) theory \cite{grossmann_lohse_2000,ahlers2009heat,stevens2013unifying}, which is 
known to capture the $Ra$ and $Pr$ dependency of $Nu$ and $Re$ over a wide parameter range. Although GL theory is based on the 
assumption that the system is three-dimensional, laterally bounded by no-slip and adiabatic walls, the agreement with our laterally periodic 
3D simulations appears satisfactory within the statistical accuracy of the numerics (see Fig. \ref{fig:comparison}a). 
Note that in all cases the Nusselt scaling exponent with $Ra$ is always smaller than $1/3$. This is better appreciated in the 
compensated plot of Fig. \ref{fig:comparison}b. 
As in \cite{van2015comparison}, we observe that the 2D RB system is less 
efficient in transporting heat than the 3D one. The highest relative difference among the 3D and 2D Nusselt numbers is of the 
order of $30\%$ and it occurs at $Ra_{eff} \approx 3 \times 10^5$ (see inset of Fig. \ref{fig:comparison}a). 
However, given the limited $Ra$-range covered, it is presently not 
possible to make statements on the variation of the scaling exponents with the Rayleigh number (see Fig. \ref{fig:comparison}b). 

Let us now turn to the CM system. An equivalent 2D-3D hierarchy is also displayed in this case. The 3D effective Nusselt 
number is always above its 2D counterpart at corresponding $Ra_{eff}$ values. In amplitude, the difference appears to be 
more important than in the corresponding RB situation, except beyond $Ra_{eff}, Ra > 10^7$ where 
it reaches similar 
values for the two systems (see inset of Fig. \ref{fig:comparison}a). 
A remarkable feature here is the enhanced heat flux displayed by the 3D CM system with respect to the RB one, 
which reaches a relative increase of $(47 \pm 6) \%$ at its maximum, 
occurring around $Ra_{eff}\simeq 5 \times 10^5$.
At larger Rayleigh numbers such a difference appears to get smaller and eventually vanish. We note, however, that this can be soundly confirmed only by performing simulations at even higher Rayleigh numbers 
($Ra_{eff}, Ra > 10^8$). 
\begin{figure}[!htb]
\begin{center}
\includegraphics[width=0.7\columnwidth]{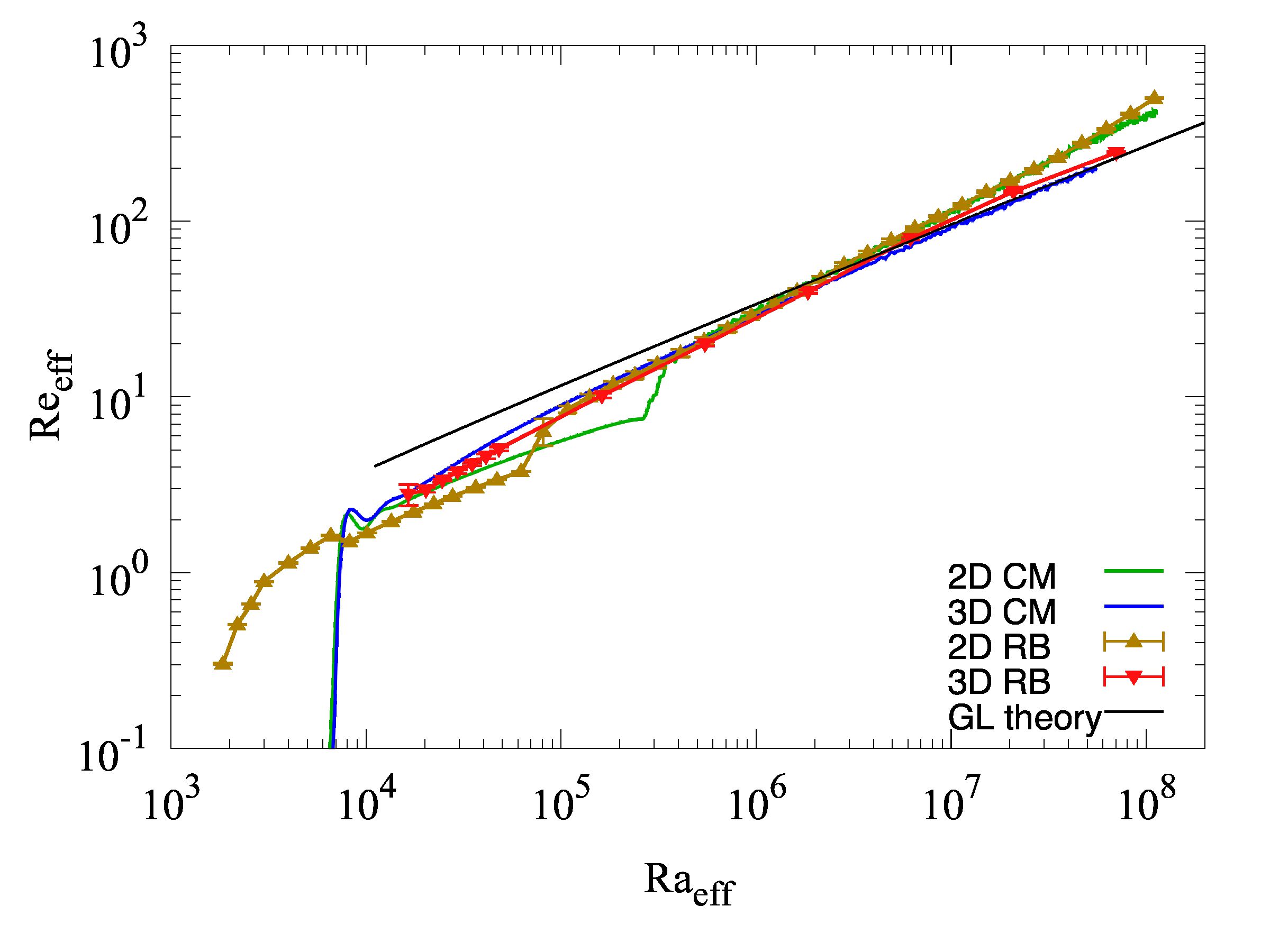}
\caption{Effective Reynolds number versus the effective Rayleigh number 
for 2D and 3D systems with the same melting process' configuration;
the global parameters are $St=1$ and $Pr=10$. The $Re \, \mathrm{vs.}  \, Ra$ data for the RB 2D and 3D cases 
are also shown. The solid black line is the prediction from
GL theory \cite{stevens2013unifying}.}
\label{fig:reynolds}
\end{center}
\end{figure}

We conclude by noting that the Reynolds number, Fig. \ref{fig:reynolds}, measured in 3D CM simulations is in nearly perfect agreement 
with both the results from 3D RB simulations and with GL theory. We observe that, contrary to the 2D case, $Re_{eff}$ in 3D 
does not show abrupt changes associated with pattern transitions. With respect to what is found for the Nusselt number, for which 
$Nu_{eff}^{3D} > Nu_{eff}^{2D}$, at high $Ra_{eff}$ we find that the effect of dimensionality is opposite for the Reynolds number, 
namely $Re_{eff}^{2D} > Re_{eff}^{3D}$,  in qualitative agreement with previous observations in 2D and 3D bounded RB systems~\cite{van2015comparison}. 

In summary, we have shown that the 3D CM system in the range of parameters studied here (and with $Pr=10, St=1$) 
behaves in a qualitatively similar manner to a RB system. The role of dimensionality is also alike in the CM and RB cases. 
However, while the Reynolds number is nearly 
identical with or without melting, the Nusselt number displays a distinct behavior characterized by $Nu_{eff}^{CM} > Nu_{eff}^{RB}$ 
for $10^4 \lesssim Ra_{eff} \lesssim 10^7$. At even higher-Ra such a difference seems to reduce and might asymptotically vanish.

\subsection{Morphology of the phase-change interface \label{sec:interface}}
In this section we aim at a quantitative characterization of the phase-change interface shape. 
The focus is on the trends as a function of the Rayleigh number at fixed Stefan number and on the possible differences 
connected with space dimensionality. With this in mind, we consider simple quantifiers of the fluid top boundary roughness 
that can be applied both in 2D and in 3D. 
\begin{figure}[!htb]
\centering
\subfigure[$Ra_{eff}=1.76\times 10^4$]{\includegraphics[width=0.325\columnwidth]{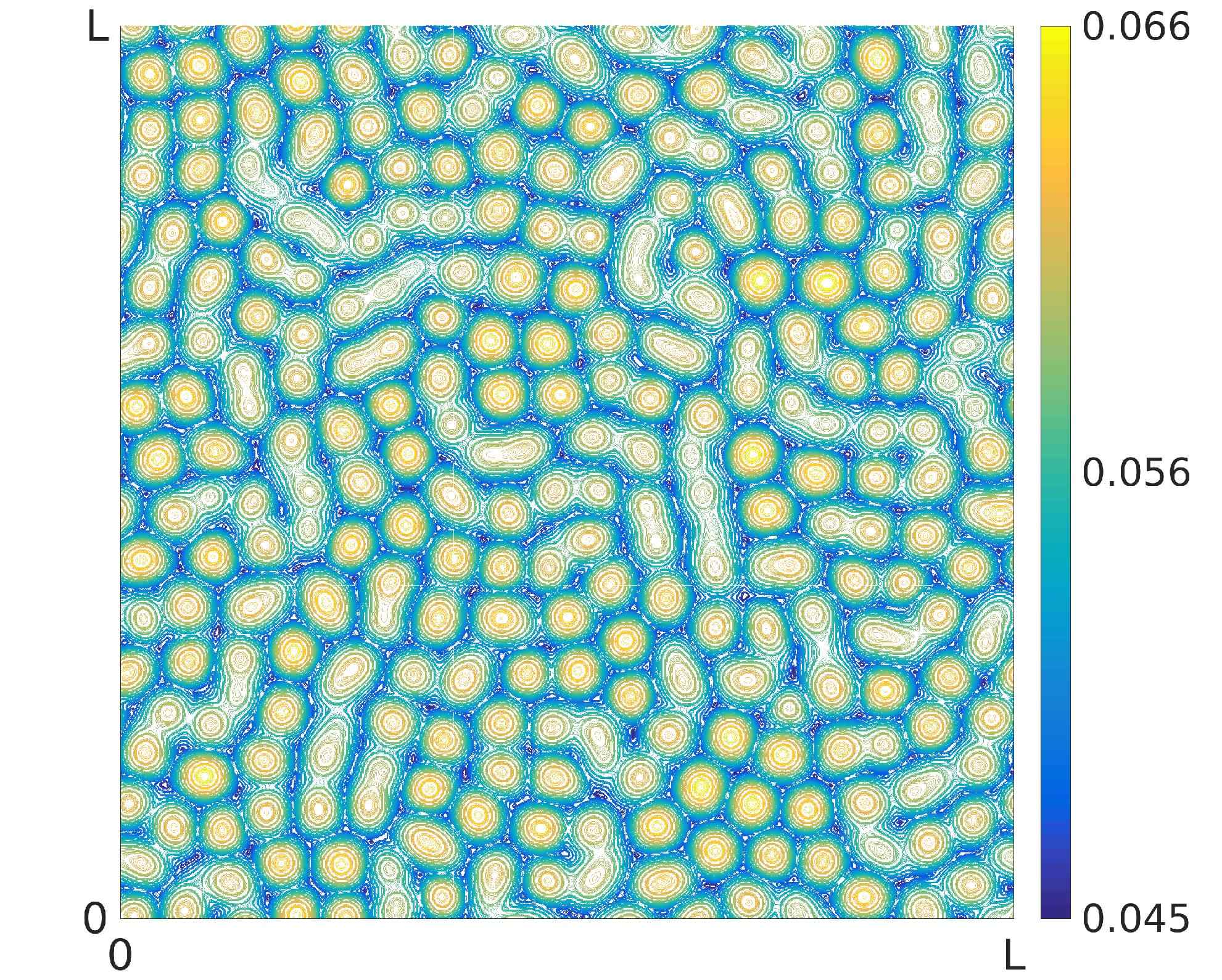}\label{fig:visualization3d-a}}
\subfigure[$Ra_{eff}=1.41\times 10^5$]{\includegraphics[width=0.325\columnwidth]{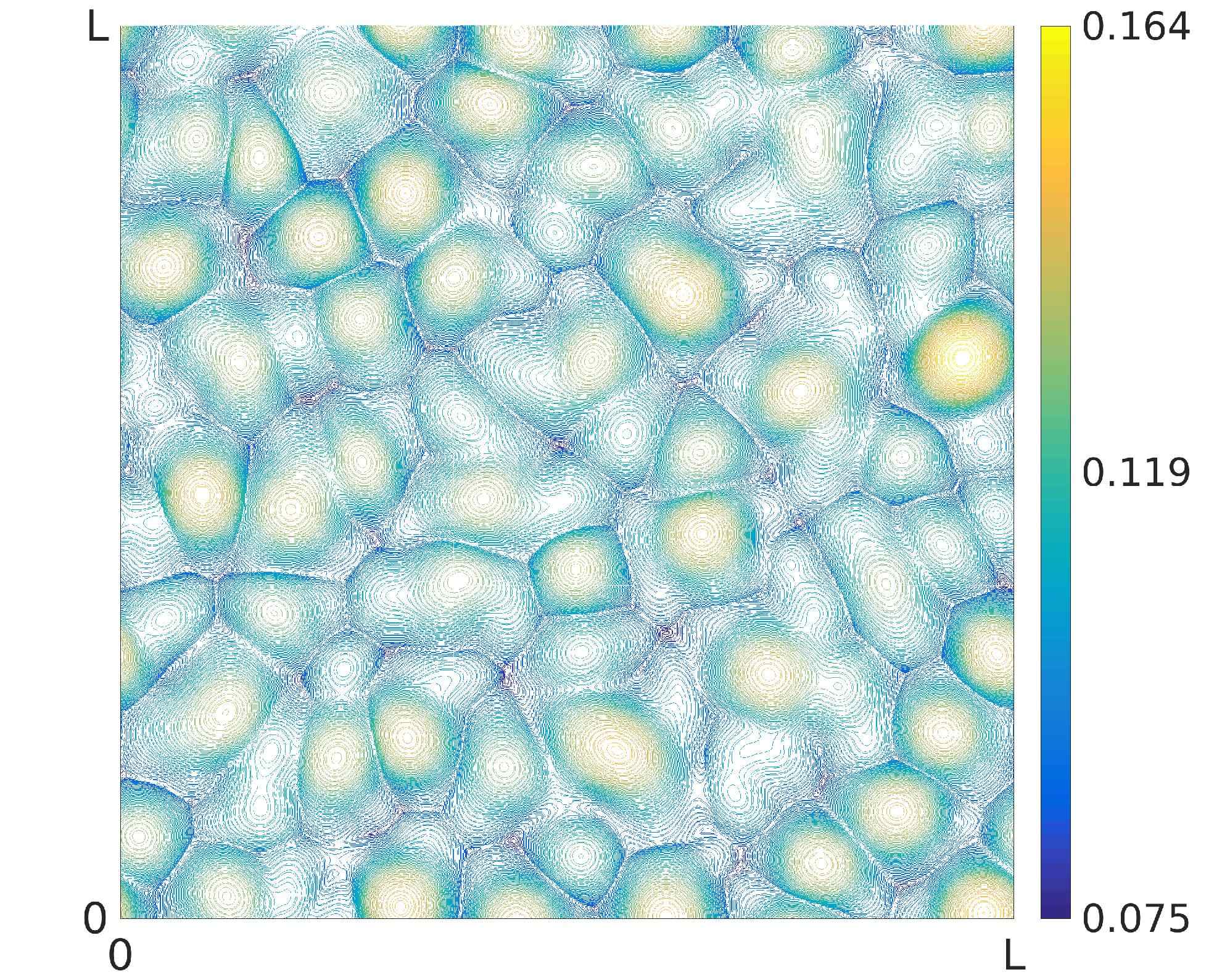}\label{fig:visualization3d-b}}
\subfigure[$Ra_{eff}=4.57\times 10^5$]{\includegraphics[width=0.325\columnwidth]{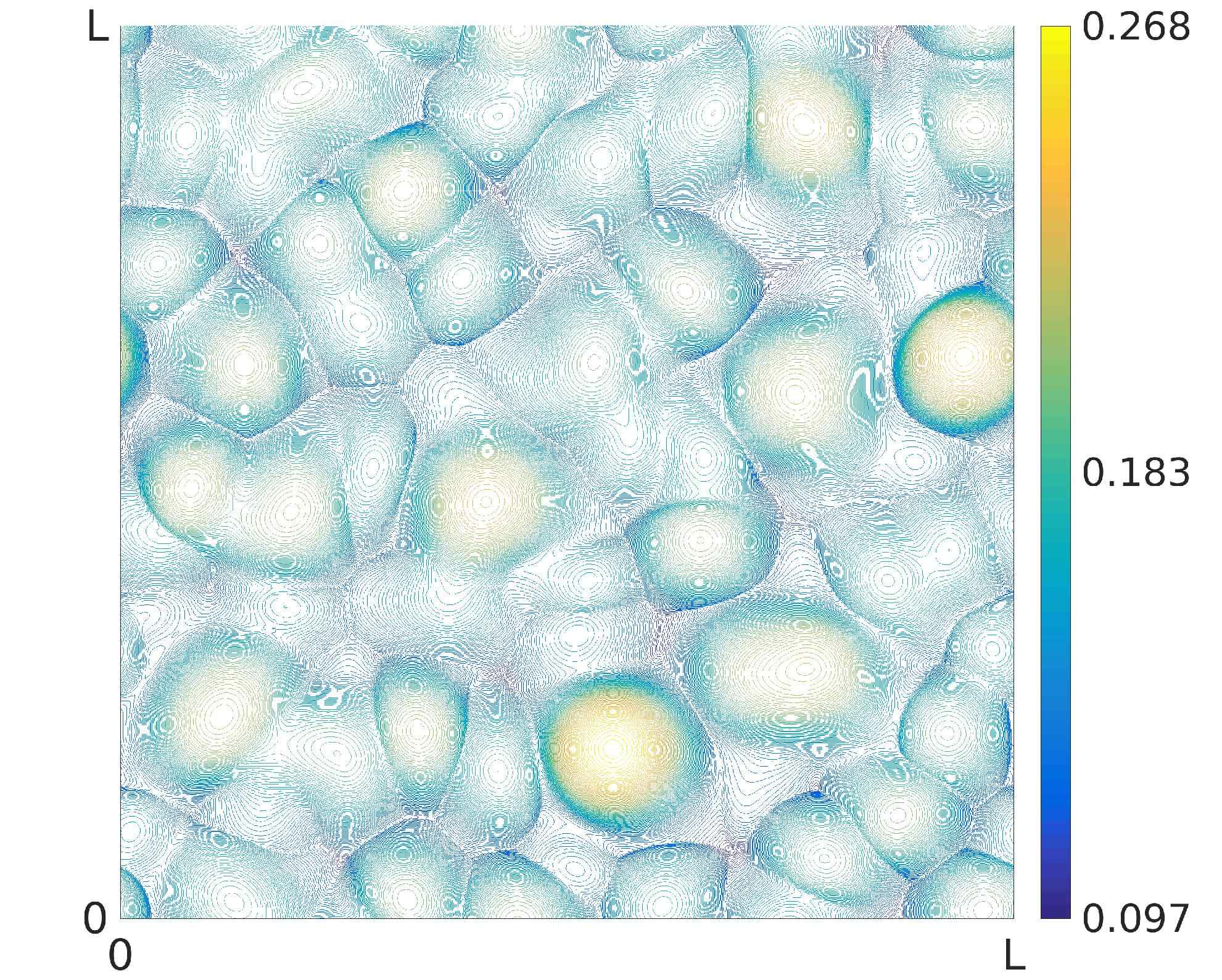}\label{fig:visualization3d-c}}
\subfigure[$Ra_{eff}=1.04\times 10^6$]{\includegraphics[width=0.325\columnwidth]{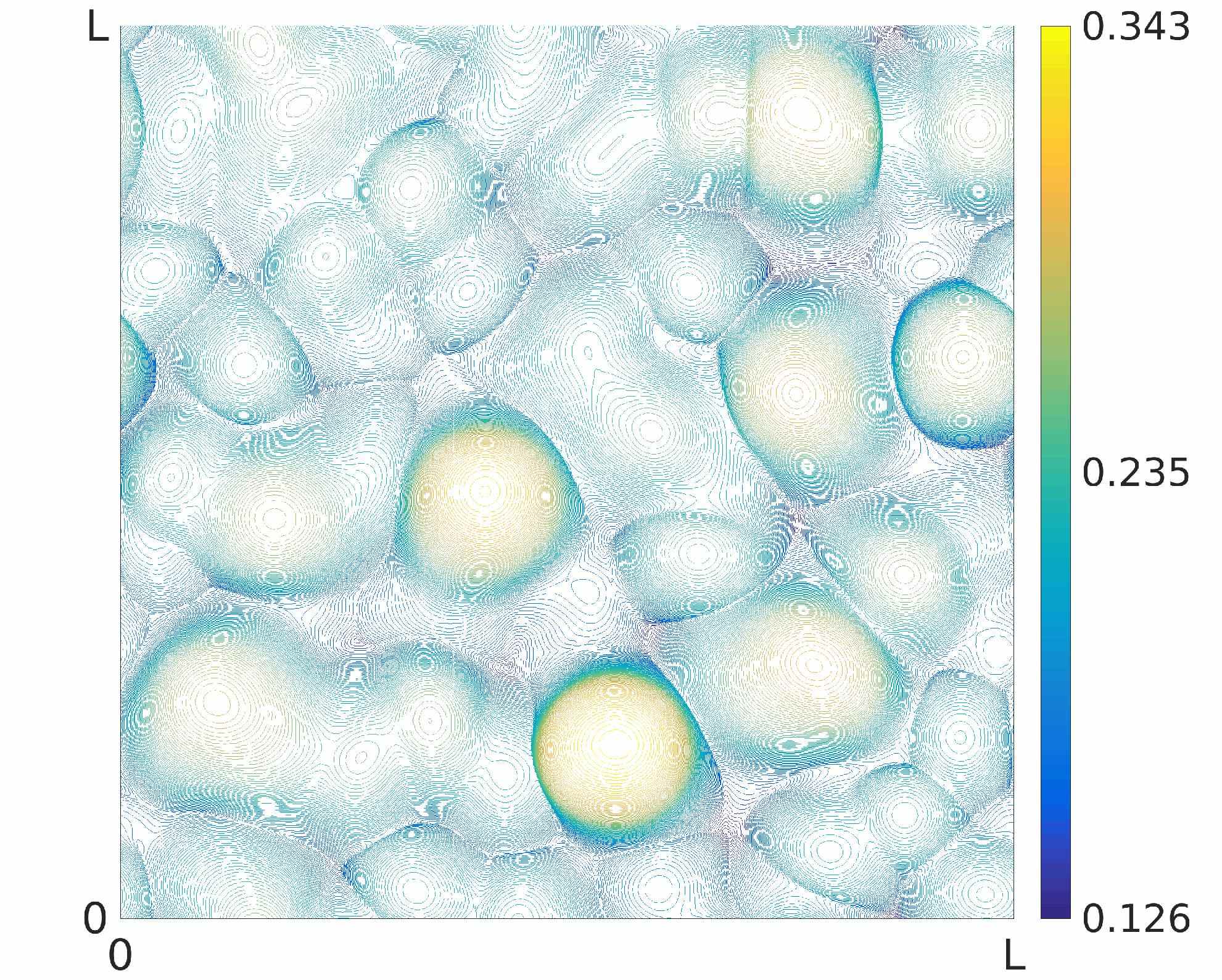}\label{fig:visualization3d-cc}}
\subfigure[$Ra_{eff}=1.15\times 10^7$]{\includegraphics[width=0.325\columnwidth]{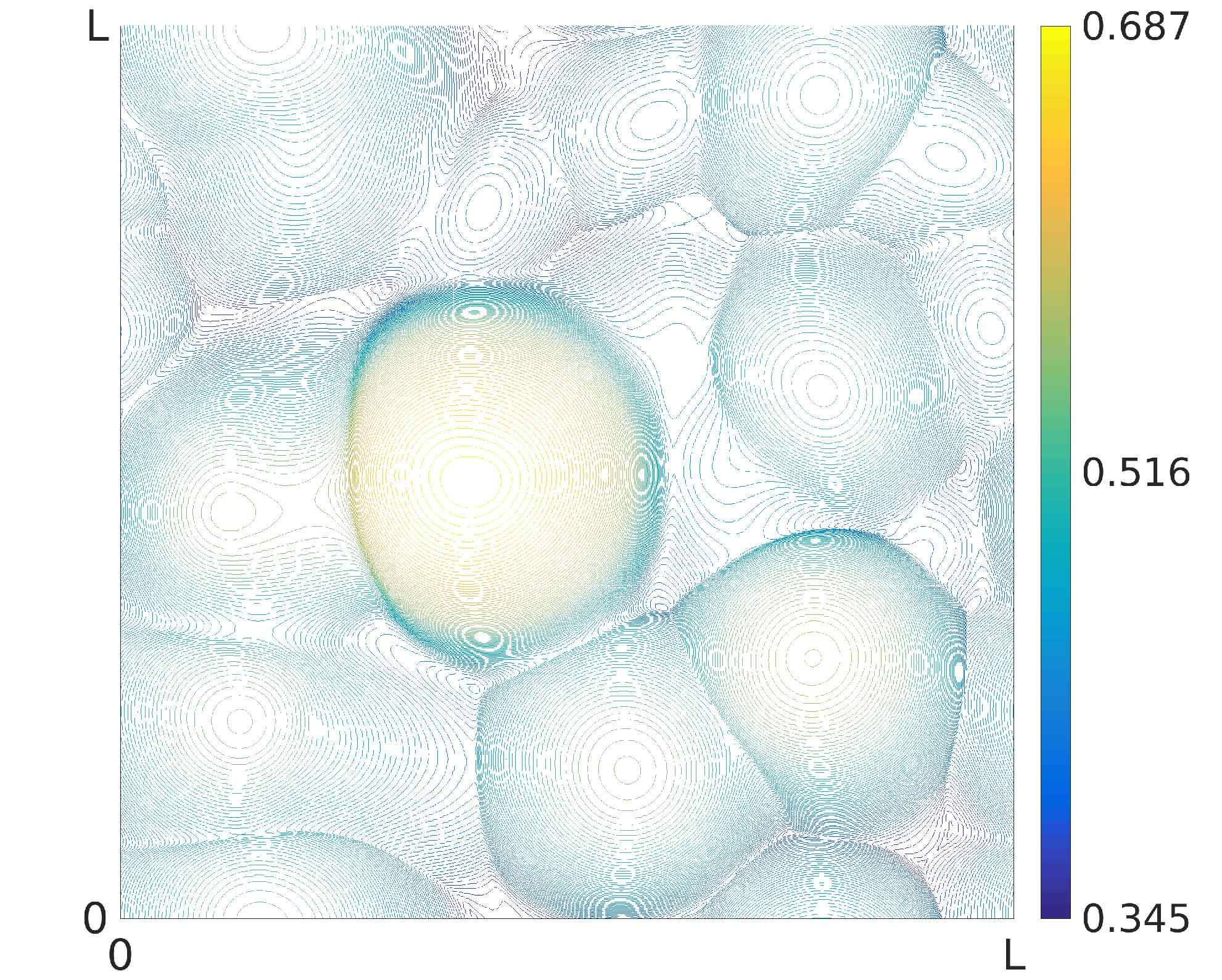}\label{fig:visualization3d-d}}
\subfigure[$Ra_{eff}=5.54\times 10^7$]{\includegraphics[width=0.32\columnwidth]{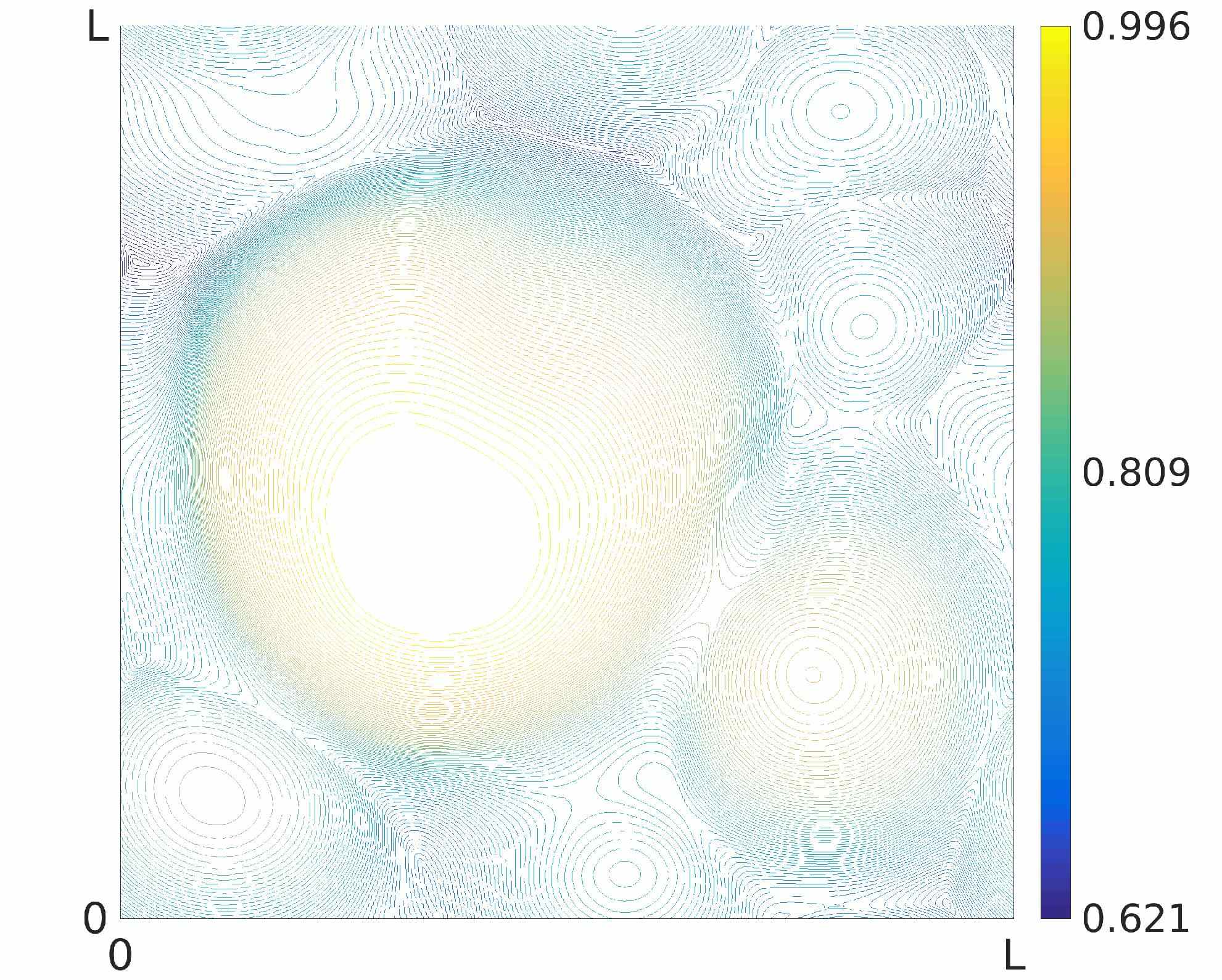}\label{fig:visualization3d-f}}
\caption{Isolines of the phase-change interface from 3D simulations at 
different Rayleigh numbers; the color codes $z_m(x,y,t)/H_{max}$. Panel (a) corresponds to a stage close to the 
convective onset where horizontally steady patterns are observed. The shape of these convective cells is found 
to be approximately hexagonal. As it can be seen, few cells have 
already merged creating elongated patterns. The merging process further intensifies at higher Rayleigh numbers (panels b-e). 
In panel (f), where the aspect ratio is $\Gamma_{eff} \simeq \Gamma_{min}=1$, a single pattern has become dominant.  
A movie of this simulation is available in the Supplemental Material (SM) online \cite{SM}.} 
\label{fig:visualization3d}
\end{figure}

Let us first discuss the phase-change surface in the 3D setup.  We will then contrast this case with the 2D one.
A sample visualization is reported in Fig. \ref{fig:visualization3d}. 
In spite of the fact that convection starts at the same value of $Ra_{eff}$ in both 2D and 3D (see Fig. \ref{fig:comparison} 
and Fig. \ref{fig:reynolds}), 
we can see from Fig. \ref{fig:visualization3d}a that, in 3D, already at $Ra_{eff}=1.76\times 10^4$ cellular-like (rather than 
roll-like) patterns form. 
This highlights the role of dimensionality. The 3D CM system displays transient polygonal patterns that subsequently merge 
into larger convective cells of similar shape (Fig. \ref{fig:visualization3d}c). This process goes on until a single big cell, 
only limited by the lateral domain size, forms.
Our results qualitatively agree with those from previous experimental investigations. Indeed, polygonal patterns of phase-change 
interfaces have already been observed in CM experiments using different substances
\cite{davis1984pattern,sugawara2007visual,hill1996}. In particular, in \cite{sugawara2007visual} cellular polygonal patterns 
were detected in the melting interface of an ice block submitted to horizontally uniform heating at its top.  

\begin{figure}[!htb]
\begin{center}
\subfigure[]{\includegraphics[width=0.493\columnwidth]{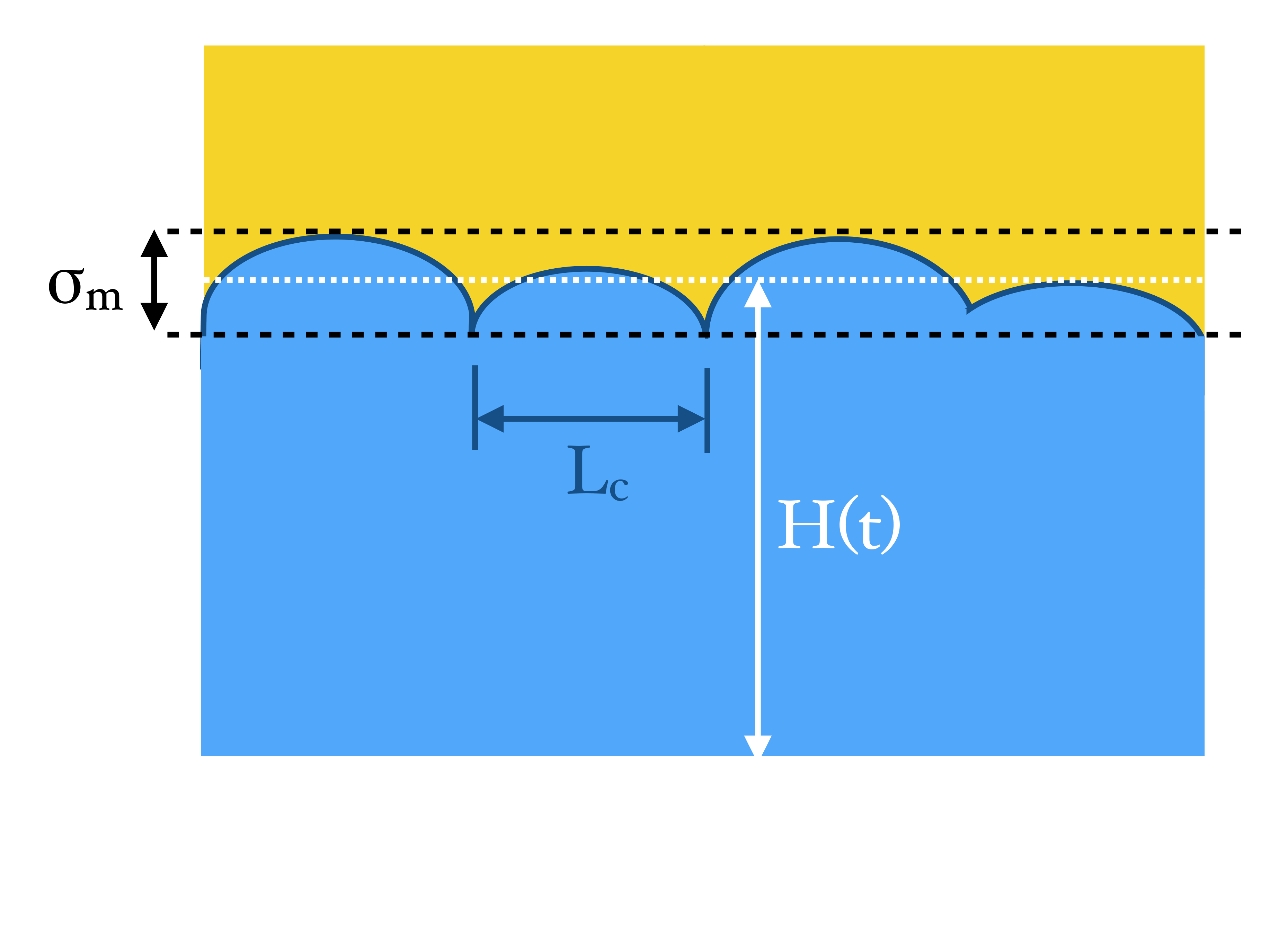}}
\subfigure[]{\includegraphics[width=0.493\columnwidth]{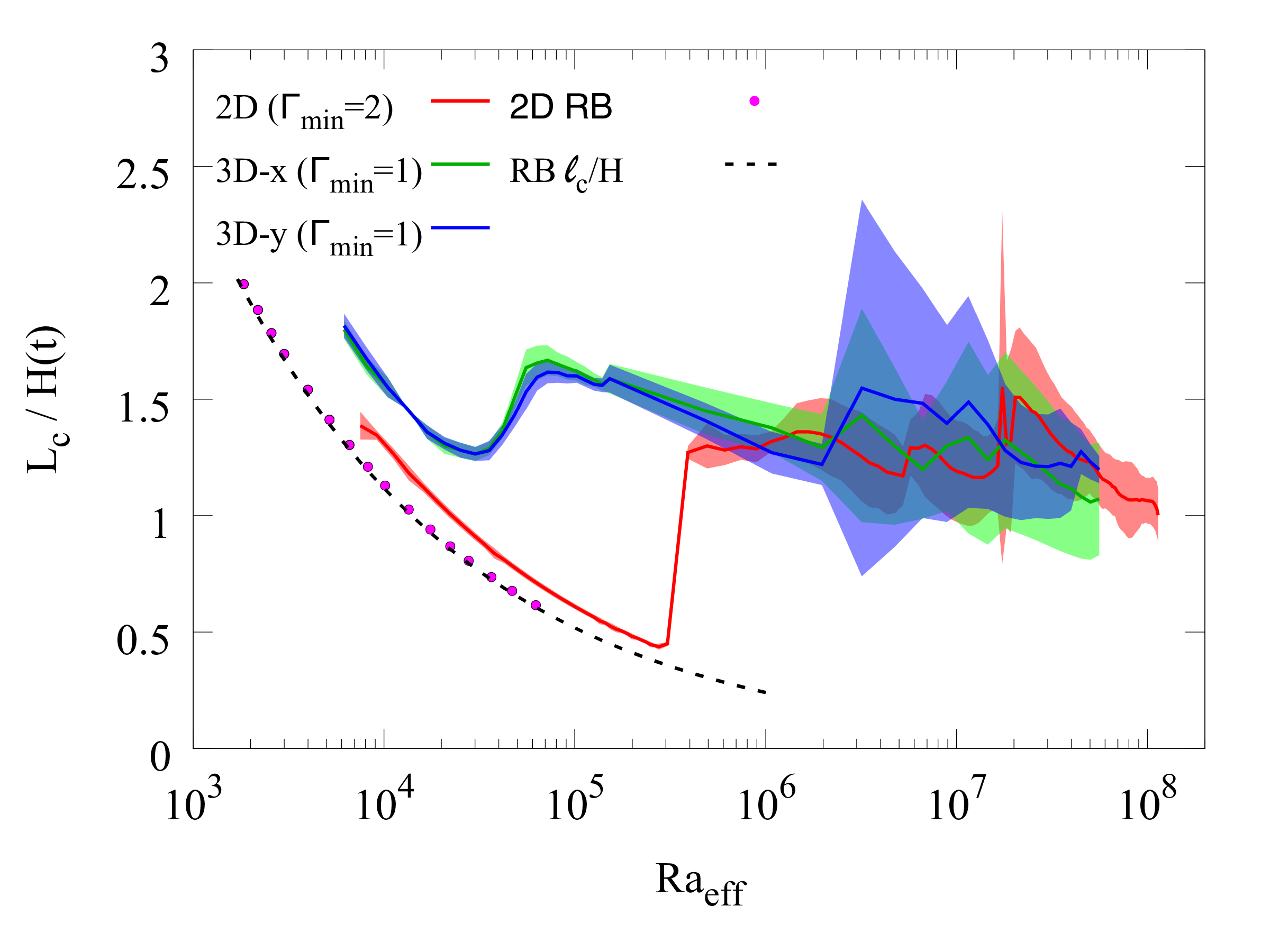} \label{fig:correlation2d3d}}
\subfigure[]{\includegraphics[width=0.493\columnwidth]{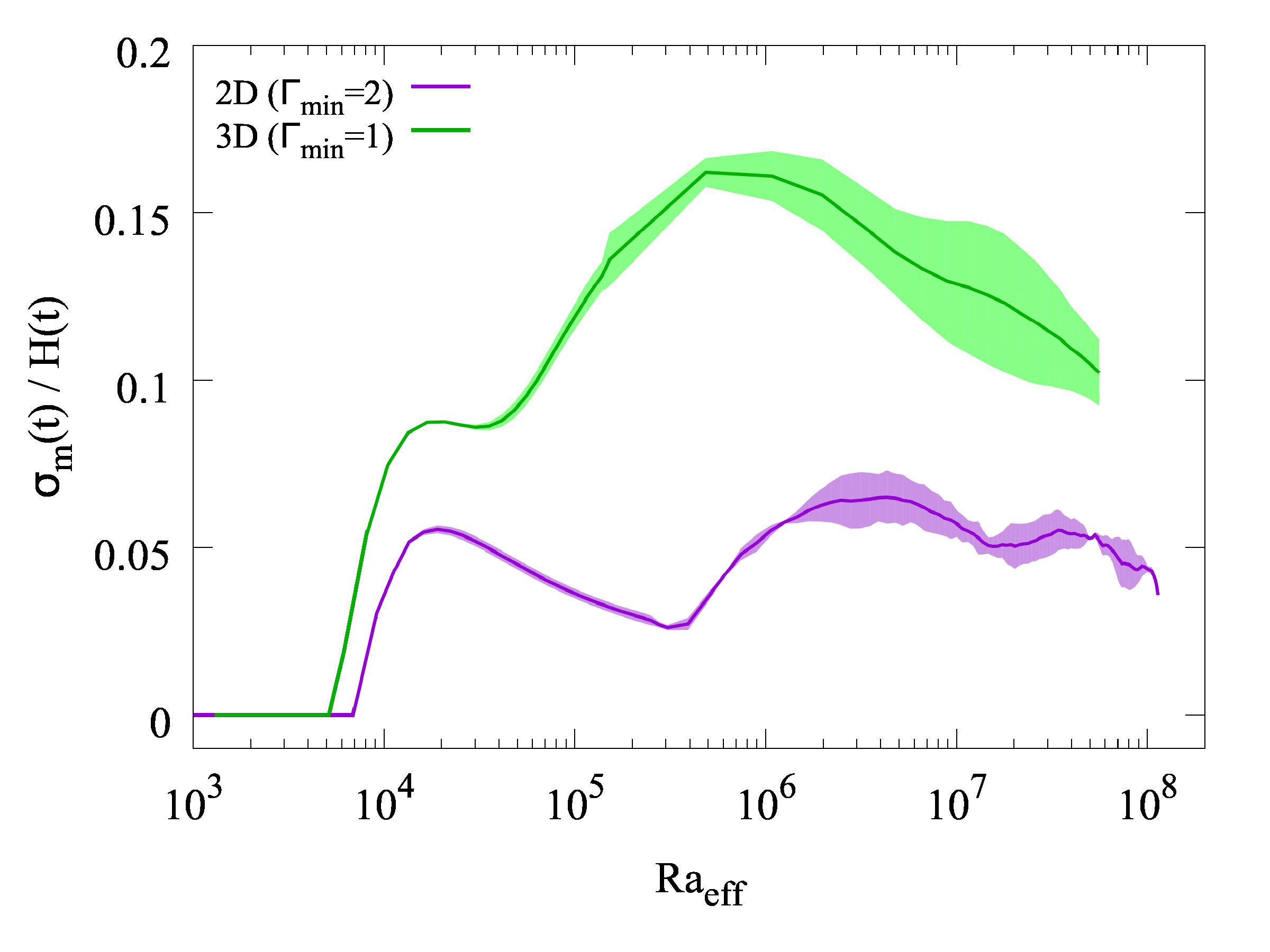}\label{fig:deviation2d3d}}
\subfigure[]{\includegraphics[width=0.493\columnwidth]{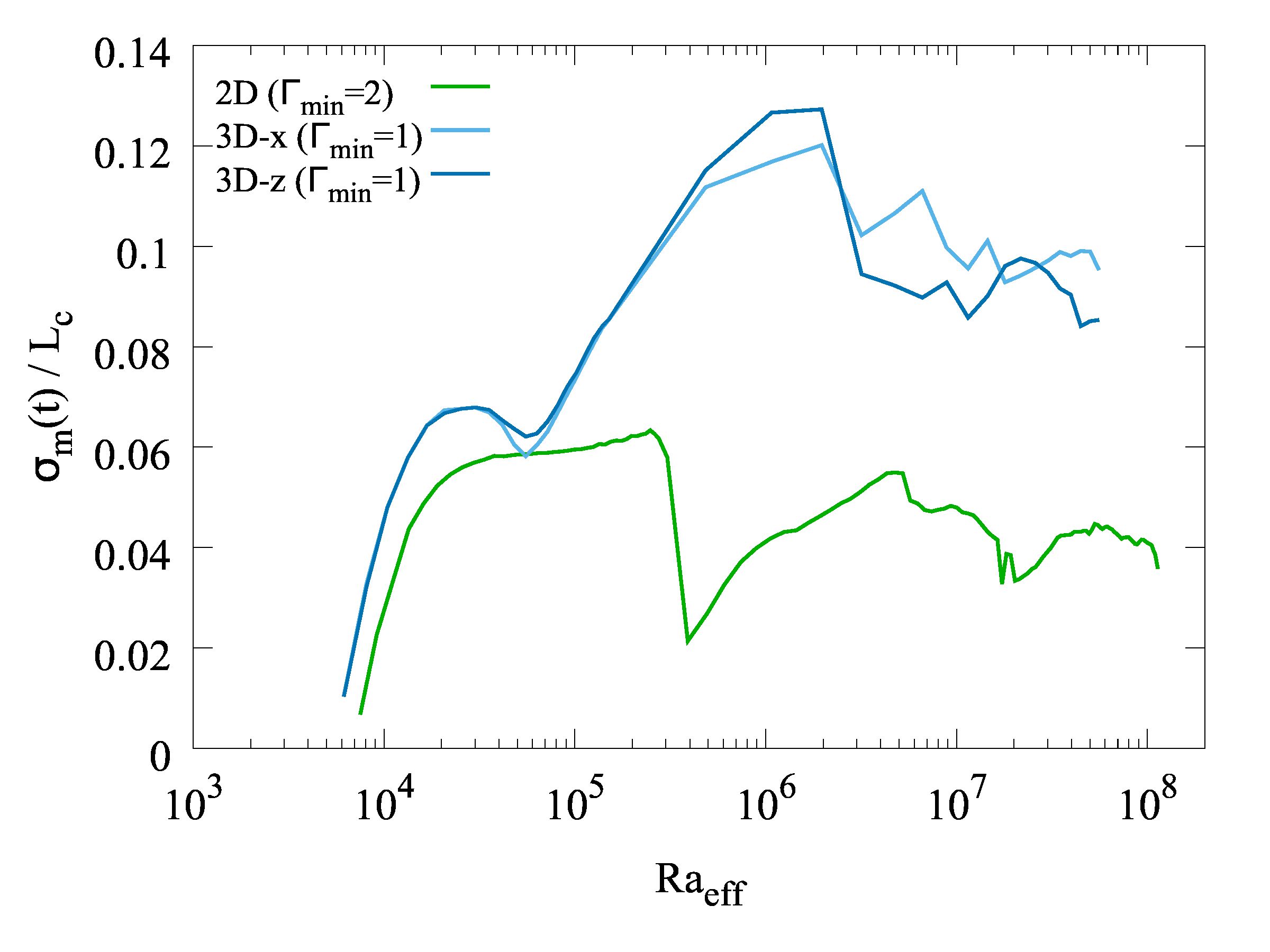}\label{fig:waveaspect2d3d}}
\caption{(a) Schematic view of the melt layer with relevant interface length scales: horizontal correlation length $L_c(t)$, 
roughness $\sigma_m(t)$, mean height $H(t)$.
(b) Correlation length versus $Ra_{eff}$. In all different cases $L_c(t)$ is at most slightly larger than one half  
of the domain width $L$ (at the end of the simulations). For comparison we also show the measurement of the lateral wavelength normalized by the 
height $H$ for the RB system, and its theoretical prediction $\ell_c/H = (2 \pi / k_c) \ (Ra_c / Ra_{eff})^{1/3}$.
(c) Roughness versus $Ra_{eff}$. (d) Ratio of the roughness to the correlation length, $\sigma_m/L_c$.
In panels (b-d), the curves are obtained from ensemble averages; the shaded areas in (b) and (c) account for the spreading 
of the measured values over different realizations, computed as the difference between the maximum and minimum values. The large 
spreading at large $Ra_{eff}$ in panel (b) is due to limited statistics.} 
\label{fig:cor_dev}
\end{center}
\end{figure}

Both in 2D and in 3D visualizations, 
the interface shape at a given time appears to be fairly well
characterized by two length scales: a single wavelength corresponding to the typical lateral size of convective patterns 
(note that in 3D these are essentially isotropic on the horizontal) and a typical roughness associated with the interface modulation 
along the vertical (see Fig. \ref{fig:cor_dev}a for a schematic view). 
To measure the first of these scales, that we call $L_c(t)$, we make use of one-dimensional auto-correlation 
functions of the local interface height. In 2D, for $ z_m(x,t)$ we have 
\begin{equation}
C(r,t) =  \langle z_m(x+r,t) z_m(x,t)\ \rangle_A
\label{eq:corr_2d}
\end{equation}
while in 3D, for $ z_m(x,y,t)$,
\begin{eqnarray}
C_x(r,t) & = & \langle z_m(x+r,y,t) z_m(x,y,t)\ \rangle_A \label{eq:corr_3dx} \\
C_y(r,t) & = & \langle z_m(x,y+r,t)  z_m(x,y,t)\ \rangle_A \label{eq:corr_3dy}
\end{eqnarray}
for the $x$ and $y$ coordinates, respectively. Clearly, we expect $C_x \approx C_y$. In the above expressions, the notation 
$\langle \ldots \rangle_A$ indicates a line average over $x$ in 2D and a surface average over $x$ and $y$ in 3D. 
Let us first observe that, in the hypothetical case of a sinusoidal interface, the position of the first minimum of 
the auto-correlation function identifies the half wavelength of the interfacial curve. By analogy we define here 
the distance for which the first minimum is attained as $L_c/2$ and we identify the longitudinal correlation length 
$L_c$ with the characteristic width of convective cells. 
The computed $L_c(t)$, normalized by $H(t)$,  as a function of $Ra_{eff}$ is shown in Fig. \ref{fig:cor_dev}b. 
Initially, i.e. for small $Ra_{eff}$, the interface is flat since convection is absent; in this case the correlation length 
is not really defined ($L_c \to \infty$). Later on, a finite $L_c$ emerges due to the onset of convection, which triggers 
the formation of recirculating patterns (cells) with an aspect ratio $\approx 1.5$. The ratio $L_c(t)/H(t)$ then decreases 
because the number of convective rolls remains constant while the height of the melt increases. 
We note that a corresponding measurement for the RB system can be performed by replacing  the interface correlation length $L_c$ by the lateral wavelength $\ell_c$ of the convection patterns (i.e. pairs of large-scale circulation rolls) . As it can be observed (Fig. \ref{fig:cor_dev}b) also the RB data decrease at increasing $Ra$ with a similar functional form as compared to the CM system, at least in the stationary regime of convection, where the identification of the width of convection patterns is unambiguous. In such a stage an analytical prediction for the scaling of $\ell_c/H$ with $Ra_{eff}$ can be obtained by taking as ingredients the critical dimensionless lateral wavenumber $k_c \simeq 3.114$ and the critical Rayleigh number $Ra_c \simeq 1708$  of the RB system \cite{chandrasekhar2013hydrodynamic}.
This leads to the expression $\ell_c/H = (2 \pi / k_c) \ (Ra_c / Ra_{eff})^{1/3}$, which closely matches the RB measurements. This proves that the characteristic correlation length of the interface in the CM system shows clear fingerprints of the convective patterns associated to the first supercritical instability of the RB system. 
Beyond a certain height of the system, both the RB and CM systems begin a cell coarsening process 
accompanied by lateral oscillations.
This is reflected in the sawtooth behavior of $L_c(t)/H(t)$, which is more evident for the more constrained 2D system. 
Asymptotically, rolls of typical aspect ratio 1 tend to prevail, independent of dimensionality. This feature is also common to 
RB convection between flat walls \cite{van2015comparison}.

The roughness of the liquid-solid interface can be quantified by means of the standard deviation of the local fluid-solid 
boundary height $z_m$, i.e.:
\begin{equation}
 \sigma_m(t) = \sqrt{\langle\ \left(  z_m(t) - H(t) \right )^2\  \rangle_A } 
\label{eq:roughness}
\end{equation}
where $\langle \ldots \rangle_A$ has the same meaning as in Eqs.(\ref{eq:corr_2d}-\ref{eq:corr_3dy}). 
The same quantity was studied by other authors with slightly different indicators \cite{davis1984pattern,hill1996,ulvrova2012dynamics}. 
The evolution of the normalized height fluctuation  $\sigma_m(t)/H(t)$ with $Ra_{eff}$ is shown in Fig. \ref{fig:cor_dev}c. 
After the initial conductive regime, in which $\sigma_m=0$, the roughness grows to approximately $(5-15)\%$ of the 
average melt height. In 3D it is typically larger, up to three times, than in 2D but the difference decreases when $Ra_{eff}$ 
is sufficiently large. The sawtooth trend is clearly visible for the 2D system.
Finally, in Fig. \ref{fig:cor_dev}d we report the evolution of the ratio $\sigma_m(t)/L_c(t)$.
This is found to be roughly constant (with the 3D value close to twice the 2D one) over three decades in $Ra_{eff}$ when the latter 
is large enough, meaning that the roughness increases as convective cells get larger.  
Stated differently, we can say that the shape of the interface, as characterized by the two discussed global scales $L_c$ and $\sigma_m$, 
appears to remain similar in the highly convective regime. Even if the physical mechanism responsible for this feature could not be 
identified and clearly deserves further studies, we advance the hypothesis that the overall shape of the interface might be determined 
by the large-scale-circulation (LSC) in each convective cell. A stronger wind would indeed produce a wider convective cell (large $L_c$) 
and proportionally a more penetrative hot flow (large $\sigma_m$).

We now discuss how the shape variations of the interface can be connected to the observed differences in the global heat flux 
in the CM system with respect to their dimensionality, or in comparison to the reference RB system.
It is known that even tiny variations of the bounding geometry of a RB cell can affect the thermal and velocity 
boundary layers and, hence, have an impact on the  mean heat-flux intensity \cite{Shen96,Ciliberto99}. 
In a variety of configurations, the Nusselt number results to be increased (see, e.g., \cite{Stringano2006} for a numerical study 
and \cite{Chilla} for a recent review).
Recently, \cite{toppaladoddi2015tailoring} have systematically 
investigated the effect of a sinusoidal top wall considering several wavelengths ($\lambda$) and a roughness ($h$)  relative to the total cell height ($H$) of $h/H= 1/10$. 
A wavelength value that is about $1/7$ with respect to the average cell height was found to be optimal in enhancing the
$Ra$-scaling of the total heat flux. Later on, in a numerical study for sinusoidal top-and-bottom walls \cite{zhuPRL2017}  
it has been demonstrated that the enhanced scaling with respect to Rayleigh is a transient feature linked to the ratios of 
the thermal boundary-layer to roughness thickness. On the opposite, the overall increase of the heat flux is not a transient 
and it is controlled by the ratio $h/\lambda$ of roughness amplitude over wall-modulation wavelength and it is maximum for $h/\lambda =1$.
Similarly to what happens in RB systems, it is then plausible that the different geometrical properties found in 2D and 3D in the present CM case are responsible of the differences detected in the 2D and 3D heat fluxes. Furthermore, in analogy to the RB phenomenology the weak heat-flux enhancement observed for CM systems as compared to RB might be 
related to the soft shape modulation of the top interface. We indeed can estimate, $h/\lambda \sim \sigma_m/L_c \simeq 0.05$ for 2D and $h/\lambda \sim0.1$ for the 3D CM system, and observe that for comparable $h/\lambda$ values rough-wall RB systems display only tiny increases of Nusselt as compared to the flat wall case \cite{zhuPRL2017}. 

\subsection{Stefan dependency \label{sec:stdependency}}
\begin{figure}[!htb]
\centering
\subfigure[]{\includegraphics[width=0.49\columnwidth]{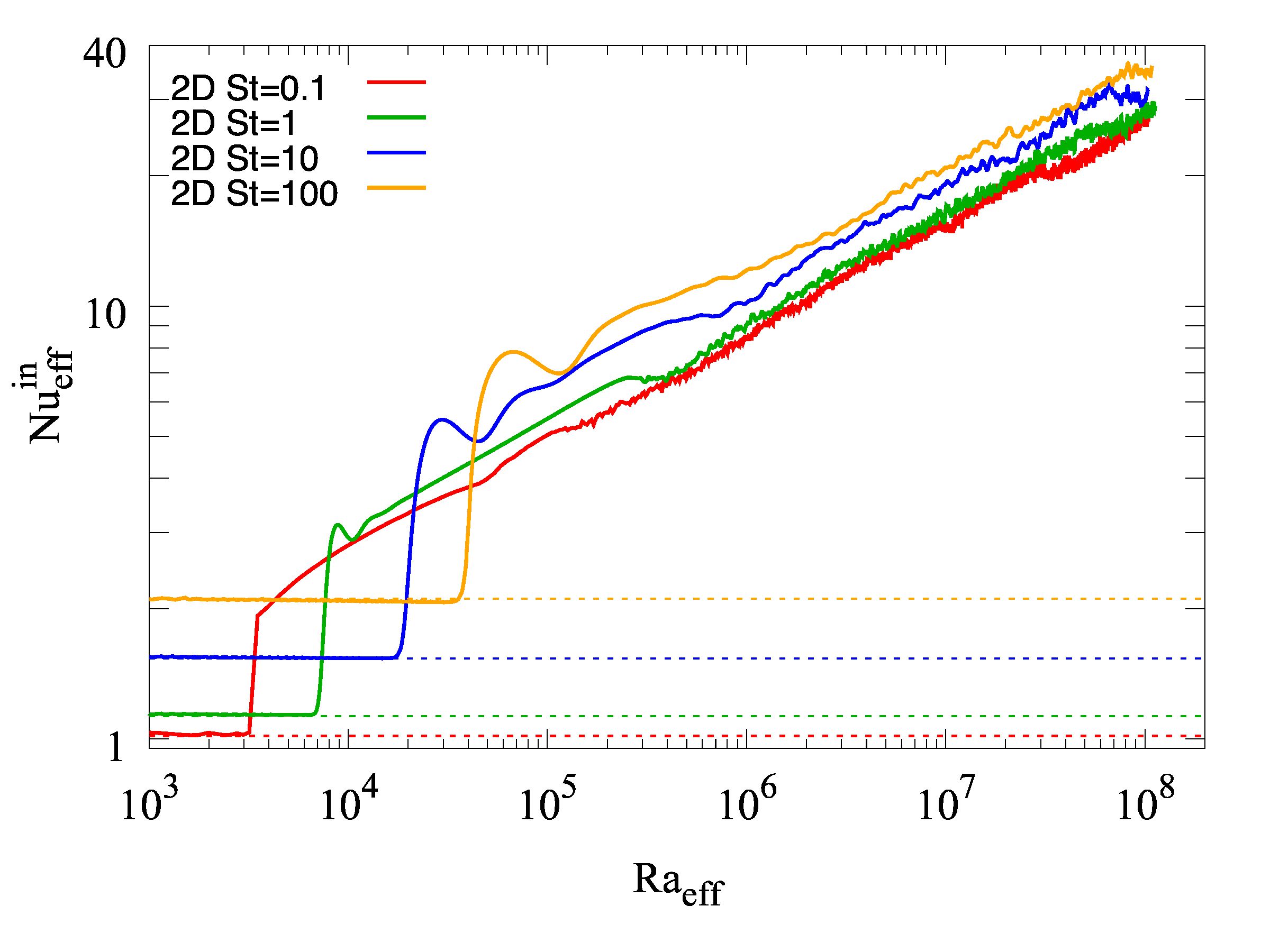}}
\subfigure[]{\includegraphics[width=0.49\columnwidth]{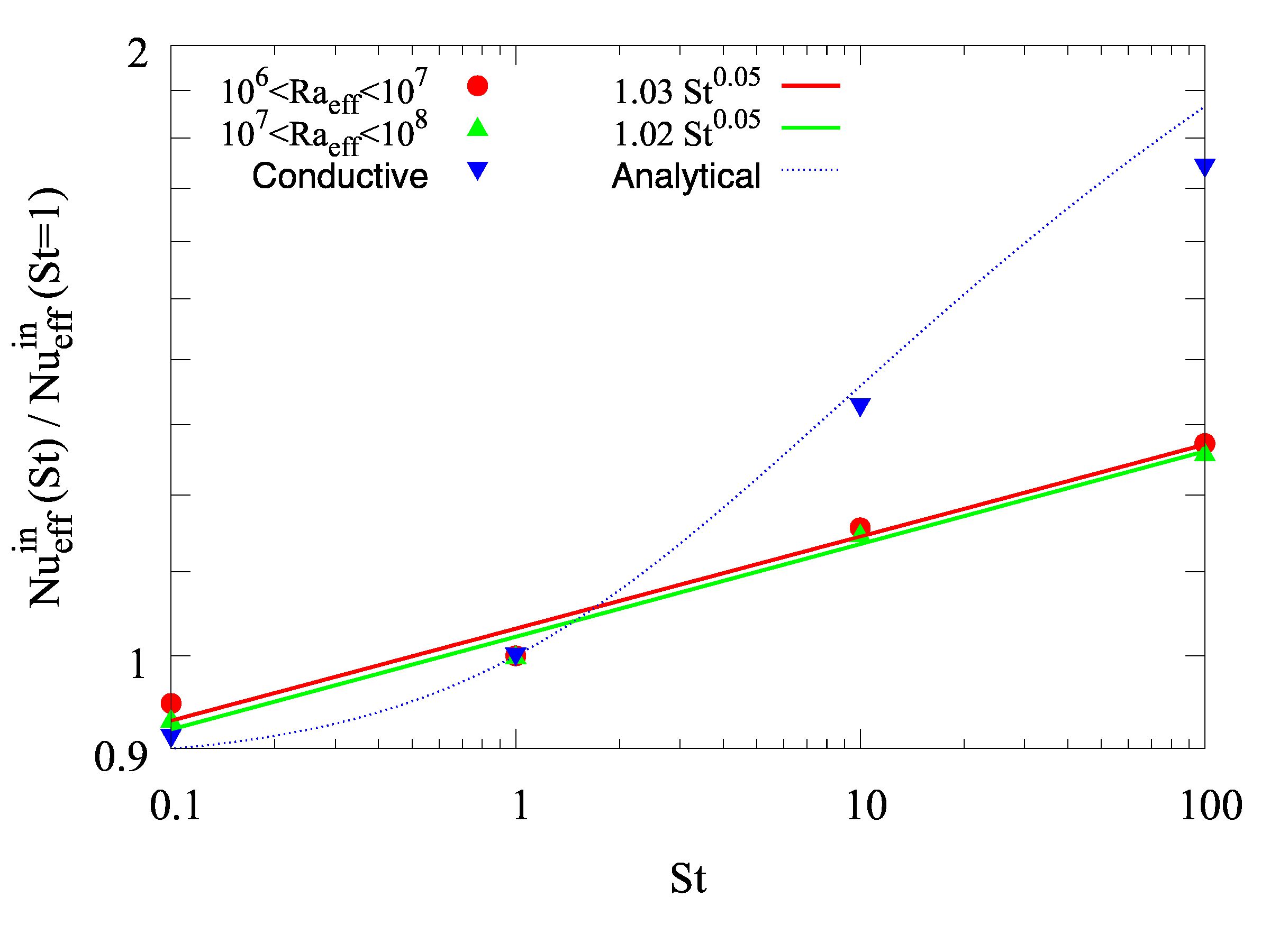}}
\caption{(a) Incoming Nusselt number versus the effective Rayleigh  at  $St=0.1,1,10,100$ in 2D. 
The horizontal dashed lines correspond to the conductive values (Eq. (\ref{eq:heatflux_conduction})). 
(b) Behavior of the 2D $Nu_{eff}^{in}$, rescaled with its value at $St=1$, for the same values of $St$ as in (a), for 
the conductive regime as well as for the convective regime ranges $Ra_{eff} \in \left[10^6,10^7 \right]$ and 
$Ra_{eff}\in\left[10^7,10^8\right]$. The analytical Stefan solution and a power law of exponent $0.05$ respectively 
account for the $St$ dependencies in the conductive and convective regimes.} 
\label{fig:stdependency1}
\end{figure}

Here we focus on the effect of varying the Stefan number on both global and morphological quantities. 
Let us first remark that high values of $St$ characterize materials for which melting is energetically inexpensive while 
low $St$ means that melting requires larger energy supply. 
In agreement with stability analysis results \cite{Kim2008}, our simulations indicate that convection arises later 
for larger $St$ (Fig. \ref{fig:stdependency1}a). If this might seem counterintuitive, it should also be noticed that the higher $St$, the larger 
the average melting front speed $v_m$. 
Moreover, figure \ref{fig:stdependency1}a shows that, in 2D, increasing $St$ causes a small but detectable increase of the incoming 
effective heat flux but apparently does not change the scaling with $Ra_{eff}$. This is better appreciated inspecting 
how $Nu_{eff}^{in}$, normalized by its reference value at $St=1$, depends on $St$ (Fig. \ref{fig:stdependency1}b). In the conductive 
regime, the growth of the rescaled $Nu_{eff}^{in}$ with $St$ is well captured by the analytical Stefan solution. After the onset 
of convection, it can be described by a power-law behavior; a best fit provides an exponent close to $0.05$ and a prefactor equal to $1$ 
within one percent precision.  In 3D, numerical simulations at $St=0.1,1,10$ (results not shown) confirm the same picture. 
Conversely, we do not detect any $St$ dependence for the scaling of the global kinetic energy. In our 2D simulations, in the range 
$St \in [0.1,100]$ the measured value of $Re_{eff}$  is essentially the same for all values of $Ra_{eff} \gtrsim 10^{6}$.

\begin{figure}[!htb]
\begin{center}
\subfigure[]{\includegraphics[width=0.495\columnwidth]{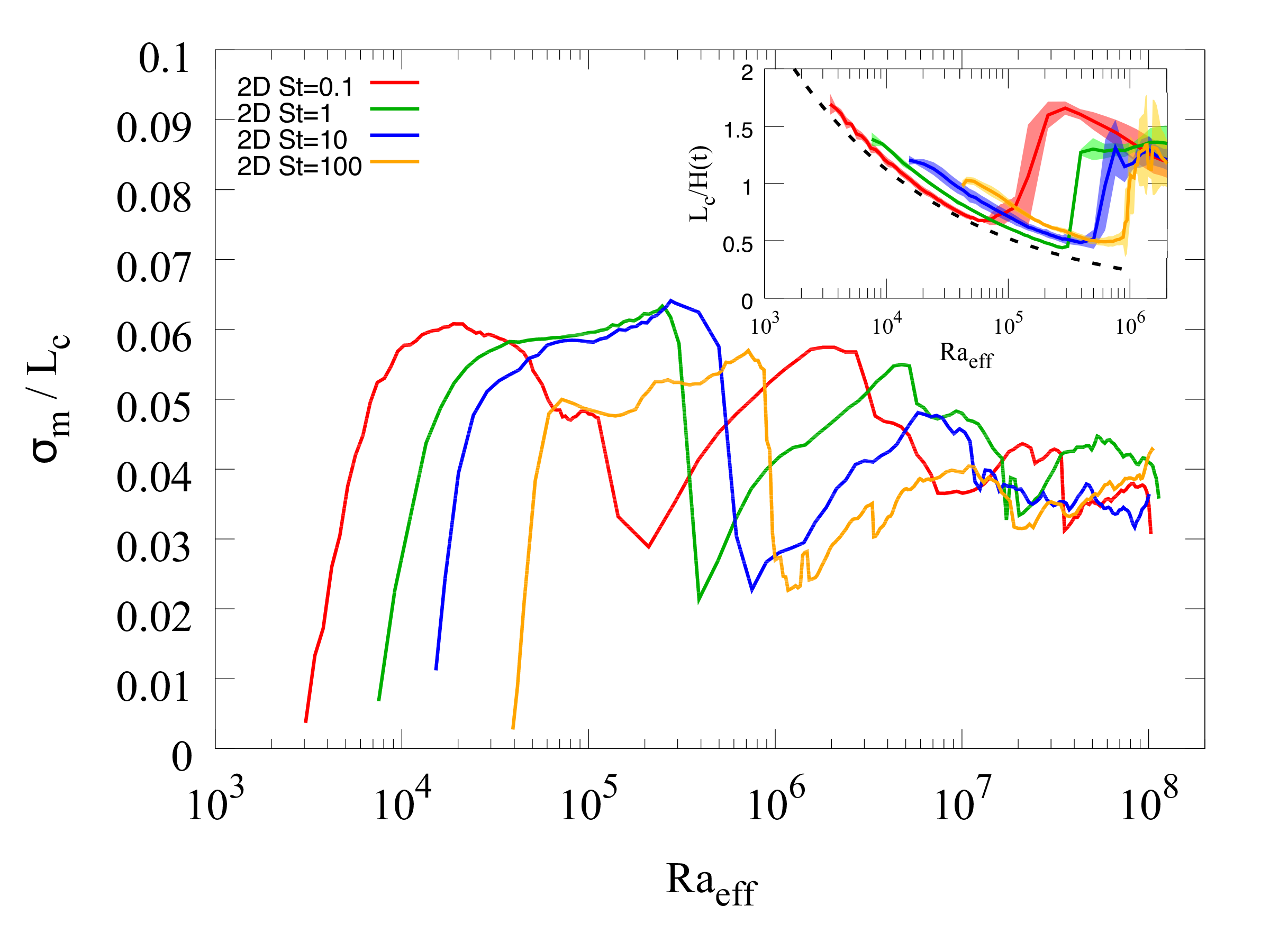}} 
\subfigure[]{\includegraphics[width=0.495\columnwidth]{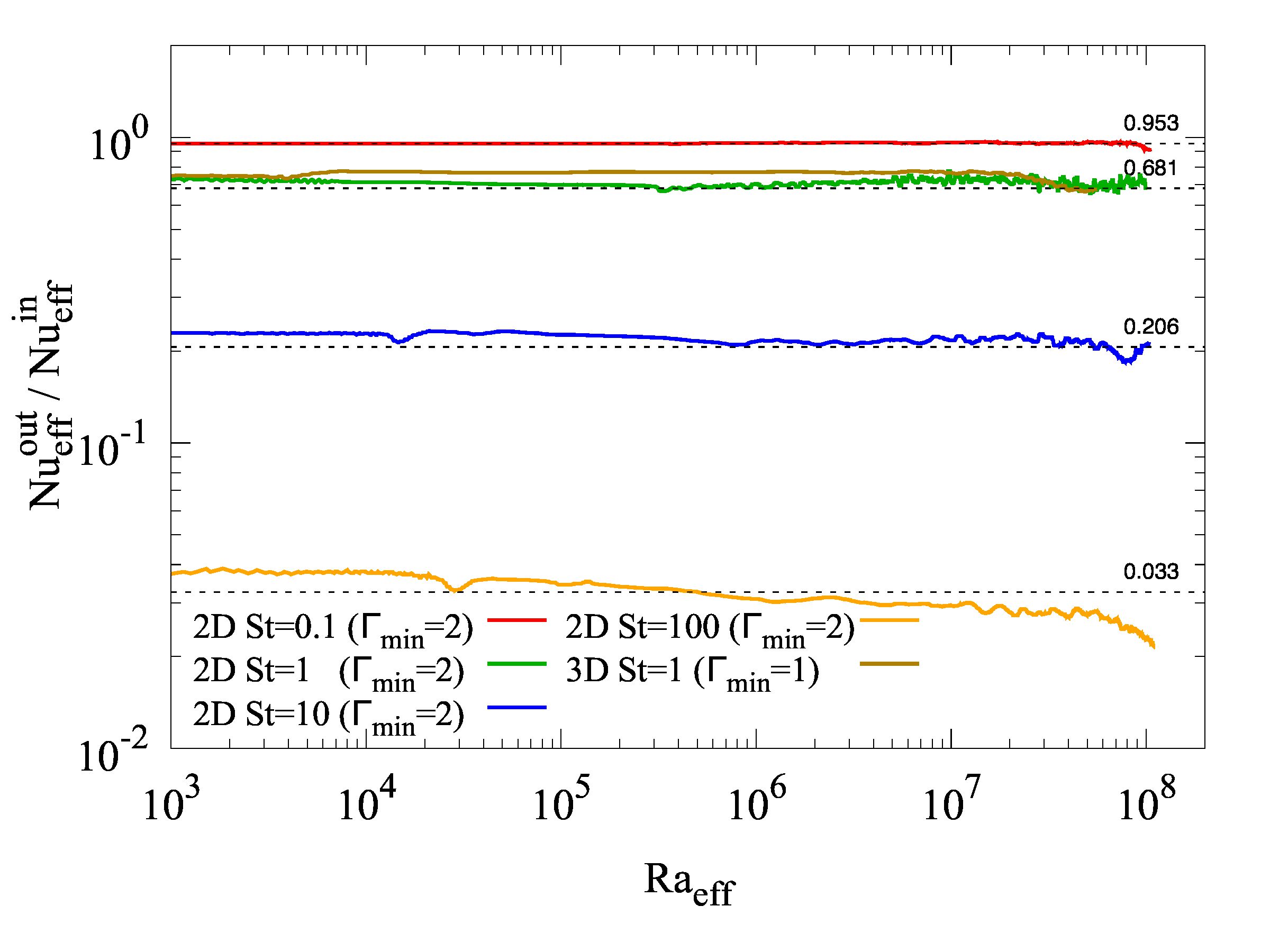}}
\caption{(a) Ratio of roughness to correlation length $\sigma_m/L_c$ of the melting interface versus $Ra_{eff}$ for $St=0.1,1,10,100$ 
in 2D. The inset shows the normalized correlation length $L_c/H(t)$, same as in 
Fig. \ref{fig:cor_dev}b, for the same Stefan values as in the main panel, 
as well as the corresponding prediction 
$(2 \pi / k_c) (Ra_c/Ra_{eff})^{1/3}$ for the RB system of equivalent height.
(b) Outgoing over incoming effective heat flux for $St=0.1,1,10,100$. 
The dashed lines are the corresponding values in conductive conditions $\exp(-\lambda^2)$. 
The 3D case at $St=1$ is also shown.}
\label{fig:fluxinout}
\end{center}
\end{figure}

We now shortly examine melting interfaces, again mainly discussing the 2D results. The horizontal correlation 
length and roughness trends confirm that the $St$ effect is weak. 
However, it is worth noting that $L_c$ decreases at smaller $St$ and progressively approaches the analytical prediction for 
the RB system close to the convective onset, see inset of Fig. \ref{fig:fluxinout}a.
It is especially interesting to examine the behavior of $\sigma_m/L_c$, because as we already discussed this ratio is 
equivalent to the ratio $h/\lambda$ of roughness amplitude over wavelength used to investigate the effect of non-flat boundaries 
in RB convection \cite{toppaladoddi2015tailoring,zhuPRL2017}. It was pointed out that the heat-flux enhancement due to wall roughness 
has a unique global maximum for $h/\lambda \simeq 1$, but then quickly decreases to the smooth-wall intensity for different $h/\lambda$ 
values \cite{zhuPRL2017}. In the present case, the always small $\sigma_m/L_c \simeq O(10^{-2})$ further decreases at increasing 
$Ra_{eff}$ and possibly attains an asymptotic plateau, (see Fig. \ref{fig:fluxinout}a), implying a nearly flat interface and 
corresponding weak variations in the heat flux as compared to a standard RB system. On the other hand, the independence of the shape 
on $St$ may be seen as a first partial confirmation to the hypothesis that the shape of the interface is related to the LSC intensity, 
and so to $Re_{eff}$, because the latter quantity is also found here to be independent of Stefan.

Finally we address the question of the inequality between the heat fluxes $Nu_{eff}^{in}$ and $Nu_{eff}^{out}$, which is tied to the 
non-stationarity of the CM system. Differently from the statistically stationary RB convection, in the CM system a fraction of the 
incoming heat is used to raise the global temperature of the cold fluid released by the melting process.
One can expect that $Nu_{eff}^{out}/ Nu_{eff}^{in}$ will be small in systems where melting is rapid, 
because most of the input heat will end up to warm the fluid. We remark that already in the conductive melting regime, 
the incoming and outgoing heat fluxes are different; the imbalance can be analytically computed as 
$Nu_{eff}^{out}/Nu_{eff}^{in} = \exp(-\lambda^2)$. 
The fraction of transmitted heat $Nu_{eff}^{out}/Nu_{eff}^{in}$ as a function of $Ra_{eff}$ is shown  in Fig. \ref{fig:fluxinout} 
for both 2D and 3D simulations at different Stefan numbers. As expected, the heat flux ratio decreases with increasing $St$, 
i.e. for faster melting. 
Interestingly, the convective values are always close to the corresponding conductive ones even when convection is very intense. 
A possible explanation of this fact might be that the heat flux ratio is controlled by the thickness of thermal boundary layers, where 
only conduction plays a role. Due to the high value of the Prandtl number used here, the thermal boundary layer is thinner than the 
velocity one and, hence, less affected by the flow field than at smaller $Pr$, which could explain the negligible impact of convection. 
However we cannot rule out the possibility that $Nu_{eff}^{in}$ and $Nu_{eff}^{out}$ depend on convection in the same way, thus giving 
a constant ratio as a function of $Ra_{eff}$ due to a compensation effect.

Let us conclude by mentioning that the present results about the link between the incoming and outgoing heat fluxes 
have a relevance for the modeling of Arctic ice melt ponds. The Stefan number of the latter being $St \approx 10^{-2}$, an extrapolation 
from our results would then indicate that $Nu_{eff}^{in} \approx Nu_{eff}^{out}$ in them. This means that, at least for what concerns 
heat-transfer features, (relatively) simple RB modeling is appropriate in that case.  

\section{Conclusions \label{sec:concl}}
We examined the dynamics of the melting process of a pure solid substance horizontally heated from below under the effect 
of conduction and natural convection by means of numerical simulations. The analysis has focused on the scaling of global 
quantities like the heat flux and the kinetic energy at varying the control parameters (the effective 
Rayleigh number $Ra_{eff}$ and the Stefan number $St$), as well as on the effects linked to space dimensionality. 
We have conducted an extensive comparison with the paradigmatic Rayleigh-B\'enard system in order to gain insight 
on the possible similiarities and differences with its dynamics. 

We have shown that CM and RB systems have similar behaviors in terms of the functional dependencies of the 
(effective) Nusselt and Reynolds numbers on the (effective) Rayleigh number; a possible reason for this was identified in the low values of the melting interface speed with respect to the typical fluid velocity fluctuations.
The $Re_{eff}$ amplitudes have been found to be almost identical in the CM and RB setups. Concerning the heat flux, $Nu_{eff}$ resulted to be larger in the CM case (and particularly in 3D), but the differences tend to vanish as the convection intensity increases (or, equivalently, asymptotically in time).
Such transient heat-flux enhancement may be related to the lower bulk temperature of the CM system as compared to the RB one. This, together with the similarity of the thermal bottom boundary layer thicknesses in both systems (a consequence of the closeness of their $Re$ numbers) should imply larger temperature gradients at the bottom wall. The fact that the CM fluid layer is colder than the corresponding RB one can be understood considering that the phase change process, represented by the Stefan term in Eq. (\ref{eq:T_nondim}) 
(which is always negative), accounts for a temperature sink. We speculate that such reduced mean temperature is associated with the increased heat pumping exerted by the system. 
With respect to the role of space dimensionality, similarly to what happens in RB convection, the global heat flux is weaker in 2D than in 3D in the CM setting. 
Altogether, these findings suggest that, in turbulent conditions, RB phenomenology can provide useful to give quantitative 
predictions for CM dynamics and that this is more true for more intense turbulence. 

Visualizations of the melting front in 3D revealed the appearance of convective patterns with approximately hexagonal, and more often 
irregular polygonal, cross section. As Rayleigh increases, i.e. as the fluid layer grows, such cells undergo a coarsening process.
Investigating the morphological properties of the liquid-solid interface with statistical indicators, we found that this is 
characterised by larger roughness in 3D than in 2D, which could be related to the differences detected in the 3D and 2D heat flux behaviors. 
However, the roughness reaches at most $15\%$ in 3D (respectively $5\%$ in 2D) of the melt height and, independent of the space 
dimensionality, it further decreases at sufficiently high Rayleigh numbers. Such low values of the melting front roughness again point 
to strong similarities between the CM and the flat-wall RB systems. 

The Stefan number dependency has been mainly investigated in 2D in the range $0.1\leq St \leq100$. Although increasing $St$ 
significantly delays the onset of convection, only quite small differences were observed in the dimensionless global heat flux, 
notably for high Rayleigh numbers. With rather good accuracy, the heat flux scaling with $St$ was found to be given by 
a power law of small exponent ($0.05$) over a broad range of $Ra_{eff}$ values. This result has potentially important consequences for 
numerical approaches, because it means that it is possible to extrapolate results of high-$St$ fast simulations to 
small-$St$ conditions that would be otherwise unattainable in direct numerical simulations. 
Because for asymptotically small $St$ the CM system approaches the RB one, an interpolation procedure for the estimation of the heat flux at any small $St$ can be devised too.
Finally, we addressed the difference between the instantaneous incoming and outgoing heat fluxes, which is connected to the 
mean temporal variation of the fluid temperature. We have shown that such heat-flux imbalance is essentially controlled by 
conductive processes and that it is more pronounced in systems where the melt process is faster, meaning for larger Stefan numbers. 

The model analyzed in this study can be seen as a simple description of ice melt ponds' dynamics. In our opinion 
the present results indicate that the heat flux measured in a corresponding RB system would give a reasonable approximation 
of the one occurring in a melt pond. Indeed, after the initial phase of the melting process, controlled by conduction, 
the non-stationary character of the CM system appears to play a minor role, due to the slow motion of the liquid-solid interface. 
Moreover, for a pond, based on the estimate $St=O(10^{-2})$, the corrugation of its bottom icy wall in our model, the relative roughness of the top boundary with respect to the average liquid layer height) can be expected to be small. 
More explicitly this tells that the average depth of a pond, $H(t)$, can be safely used for the estimate of the albedo because the modulation of the bottom topography is likely to be negligible.

We remark, however, that other important factors, here ignored for simplicity, participate in the dynamics of real ice melt ponds.  
Among these, it seems to us that it would be particularly interesting to consider the thermal forcing due to solar-radiation heating and 
the effect of wind drafts at the air-water interface (the pond top boundary), as they could affect the instability and evolution of 
convective patterns. Addressing such effects needs careful investigations that are left for future work. 

\section*{Acknowledgments}
\noindent
The research leading to these results has received funding from the French National Agency for Research (ANR) under the grant SEAS (ANR-13-JS09-0010) and from European COST Action MP1305. We acknowledge N. Ouarzazi, M. Vancoppenolle, A. Scagliarini and F. Toschi for useful discussions.
\appendix
\section{Conductive melting regime \label{sec:conductive}}
The conductive solution of the melting problem, known as Stefan solution \cite{alexiades1992mathematical}, provides the evolution of 
the liquid-solid interface and of the temperature field in the absence of convection. Reasoning for simplicity on a one-dimensional 
system, one obtains: 
\begin{equation}
z_m(t)=2\lambda\sqrt{\kappa t}, \qquad \textrm{with} \qquad  \lambda\exp(\lambda^2)\erf{\lambda}=\frac{\text{St}}{\sqrt{\pi}}
\label{eq:lambda}
\end{equation}
for the height of the fluid layer, where $\lambda$ is a constant specified by the above implicit relation 
(recall that $St$ depends on $\lambda$). 
For the temperature $T_c$ of the fluid layer in this conductive state one has:
\begin{equation}
T_c (z,t)=T_{0}-(T_{0}-T_m)\frac{\erf{\lambda\ \frac{z}{z_m(t)}}}{\text{erf}(\lambda)},
\label{eq:T_stefan}
\end{equation}
where $\mathrm{erf(...)}$ is the error function.
In 2D or 3D the front associated with the phase change will stay flat (and horizontal in our setting) at all times.  
The Stefan-like solution implies 
\begin{equation}
\langle \phi_l \rangle = 2\lambda\sqrt{\tilde{t}},
\label{eq:phi_stefan}
\end{equation}
in non-dimensional units, for the liquid fraction. Plugging (\ref{eq:T_stefan}) in Eq. (\ref{eq:summarynu_in}) and (\ref{eq:phi_stefan}) 
in Eq. (\ref{eq:summarynu_out}) one obtains the incoming and outgoing heat fluxes of Eq. (\ref{eq:heatflux_conduction}).

\section{Numerical method \label{sec:numerics}}
Several computational methods are currently available for the description of problems where fluid dynamics is coupled to 
the process of solid-to-liquid phase change. Classical computational fluid-dynamics methods, namely finite-volume or finite-difference 
discretizations of the evolution equations, either based on a moving-boundary (front-tracking) or on single-domain 
(enthalpy formulation) schemes have been extensively employed. A benchmark review on the accuracy of the different numerical methods 
used to treat laterally heated CM systems can be found in \cite{Bertrand1995}. 
Basal melting in a geophysical context was more recently addressed in \cite{Ulvrova}. In that work, the performances of the 
two above mentioned approaches were also compared. In the last decade computational schemes for the treatment of the solid-to-liquid 
phase transition have been extended also to the mesoscopic LB method for fluid dynamics. Such schemes can be classified in two groups: 
(i) phase-field methods relying on Ginzburg-Landau theory (\cite{miller2002lattice,rasin2005phase,medvedev2005lattice}); 
(ii) enthalpy-based methods (\cite{jiaung2001lattice,chatterjee2005enthalpy,Huber}) which are in fact very similar to the classic ones.

In the present work, a LB algorithm based on an enthalpy formulation of the phase-change process is chosen, similarly to the one proposed 
in \cite{Huber}. The scheme makes use of a single relaxation time LB algorithm, $D2Q9$ and $D3Q19$ lattice topologies, and a multi-population 
method to resolve the fluid velocity and temperature equations, coupled to an iterative enthalpy-based method to obtain the melt fraction field. 
The phase-change term is therefore introduced as a source/sink  term in the temperature equation. 
In short, once the temperature has been calculated at the discrete time $t_n$ we proceed to the evaluation of the local enthalpy 
(i.e. for any given position in the computational domain):
\begin{equation}
\mathcal{H}(t_n)=c_p\ T(t_n)+\mathcal{L}\ \phi_l(t_{n-1}).
\label{eq:enthalpy_num}
\end{equation}
This is used to estimate the melt fraction at time $t_{n}$ through a linear interpolation:
\begin{equation}
\phi_l(t_{n})=\left\{
\begin{tabular}{l @{\hskip 0.2in} l}
0                                                                                    &$\mathcal{H}(t_n) < \mathcal{H}_s = c_pT_m$,\\
$\frac{\mathcal{H}(t_n)-\mathcal{H}_s}{\mathcal{L}}$ &$\mathcal{H}_s\le\mathcal{H}(t_n)\le\mathcal{H}_l$,\\
1                                                                                    &$\mathcal{H}(t_n) > \mathcal{H}_l = c_pT_m + \mathcal{L},$
\end{tabular}
\right.
\label{eq:phi_update_num}
\end{equation}
and finally the liquid fraction increment is estimated by a first-order finite difference
\begin{equation}
  \frac{\partial \phi_l}{\partial t}(t_n) \simeq \frac{\phi_l(t_{n}) - \phi_l(t_{n-1})}{t_{n} - t_{n-1}}. 
\label{eq:der_phi_num}
\end{equation}
Such a term is used to define a source term in the temperature equation, which is updated for computing $T(t_{n+1})$ after a propagation 
and collision step of the LB algorithm. To increase the precision of this algorithm one could repeat the above procedure iteratively, 
however it has been shown in \cite{Huber} that a single iteration is sufficient to reach good agreement with the known analytical results 
in the conductive regime. In order to avoid the possibility of deforming the solid due to spurious numerical velocities in the part of the 
domain corresponding to it, we apply the following two corrections. First, all external forces to the system are weighted proportionally 
to the local liquid fraction  $\phi_l(\bm{x},t)$; in the present case this means that the buoyancy force does not act in the solid phase. 
Second, we apply a penalization force that strongly depends on $\phi_l$: 
\begin{equation}
\bm{f}_p = - \chi(\phi_l) \bm{u}
\label{eq:mask_num}
\end{equation}
where $\chi(\phi_l)=1-\phi_l^2$ is a penalization mask. We have checked that the specific form of $\chi(\phi_l)$ does not affect the results. 
The phase-change LB algorithm with penalization method was previously introduced by \cite{chatterjee2005enthalpy}. 
Other authors, as e.g. \cite{Ulvrova}, impose a strong (i.e. exponential) dependence of viscosity on the solid fraction 
$1-\phi_l$. 
We thoroughly validated our algorithm against known solutions of the Stefan problem, as well as by comparing its results with 
other numerical results in convective melting configurations with lateral heat source \cite{Bertrand1995}.

\begin{thebibliography}{43}%
\makeatletter
\providecommand \@ifxundefined [1]{%
 \@ifx{#1\undefined}
}%
\providecommand \@ifnum [1]{%
 \ifnum #1\expandafter \@firstoftwo
 \else \expandafter \@secondoftwo
 \fi
}%
\providecommand \@ifx [1]{%
 \ifx #1\expandafter \@firstoftwo
 \else \expandafter \@secondoftwo
 \fi
}%
\providecommand \natexlab [1]{#1}%
\providecommand \enquote  [1]{``#1''}%
\providecommand \bibnamefont  [1]{#1}%
\providecommand \bibfnamefont [1]{#1}%
\providecommand \citenamefont [1]{#1}%
\providecommand \href@noop [0]{\@secondoftwo}%
\providecommand \href [0]{\begingroup \@sanitize@url \@href}%
\providecommand \@href[1]{\@@startlink{#1}\@@href}%
\providecommand \@@href[1]{\endgroup#1\@@endlink}%
\providecommand \@sanitize@url [0]{\catcode `\\12\catcode `\$12\catcode
  `\&12\catcode `\#12\catcode `\^12\catcode `\_12\catcode `\%12\relax}%
\providecommand \@@startlink[1]{}%
\providecommand \@@endlink[0]{}%
\providecommand \url  [0]{\begingroup\@sanitize@url \@url }%
\providecommand \@url [1]{\endgroup\@href {#1}{\urlprefix }}%
\providecommand \urlprefix  [0]{URL }%
\providecommand \Eprint [0]{\href }%
\providecommand \doibase [0]{http://dx.doi.org/}%
\providecommand \selectlanguage [0]{\@gobble}%
\providecommand \bibinfo  [0]{\@secondoftwo}%
\providecommand \bibfield  [0]{\@secondoftwo}%
\providecommand \translation [1]{[#1]}%
\providecommand \BibitemOpen [0]{}%
\providecommand \bibitemStop [0]{}%
\providecommand \bibitemNoStop [0]{.\EOS\space}%
\providecommand \EOS [0]{\spacefactor3000\relax}%
\providecommand \BibitemShut  [1]{\csname bibitem#1\endcsname}%
\let\auto@bib@innerbib\@empty
\bibitem [{\citenamefont {Solomatov}(2015)}]{solomatov}%
  \BibitemOpen
  \bibfield  {author} {\bibinfo {author} {\bibfnamefont {V.}~\bibnamefont
  {Solomatov}},\ }\href {\doibase
  https://doi.org/10.1016/B978-0-444-53802-4.00155-X} {\emph {\bibinfo {title}
  {Treatise on Geophysics}}},\ \bibinfo {edition} {2nd}\ ed.,\ edited by\
  \bibinfo {editor} {\bibfnamefont {Gerald}\ \bibnamefont {Schubert}}\
  (\bibinfo  {publisher} {Elsevier},\ \bibinfo {address} {Oxford},\ \bibinfo
  {year} {2015})\ pp.\ \bibinfo {pages} {81 -- 104}\BibitemShut {NoStop}%
\bibitem [{\citenamefont {Brandeis}\ and\ \citenamefont
  {Jaupart}(1986)}]{brandeis1986interaction}%
  \BibitemOpen
  \bibfield  {author} {\bibinfo {author} {\bibfnamefont {Genevi{\`e}ve}\
  \bibnamefont {Brandeis}}\ and\ \bibinfo {author} {\bibfnamefont {Claude}\
  \bibnamefont {Jaupart}},\ }\bibfield  {title} {\enquote {\bibinfo {title} {On
  the interaction between convection and crystallization in cooling magma
  chambers},}\ }\href {\doibase https://doi.org/10.1016/0012-821X(86)90145-7}
  {\bibfield  {journal} {\bibinfo  {journal} {Earth Planet. Sci. Lett.}\
  }\textbf {\bibinfo {volume} {77}},\ \bibinfo {pages} {345 -- 361} (\bibinfo
  {year} {1986})}\BibitemShut {NoStop}%
\bibitem [{\citenamefont {{Brandeis}}\ and\ \citenamefont
  {{Marsh}}(1989)}]{brandeis1989convective}%
  \BibitemOpen
  \bibfield  {author} {\bibinfo {author} {\bibfnamefont {G.}~\bibnamefont
  {{Brandeis}}}\ and\ \bibinfo {author} {\bibfnamefont {B.~D.}\ \bibnamefont
  {{Marsh}}},\ }\bibfield  {title} {\enquote {\bibinfo {title} {{The convective
  liquidus in a solidifying magma chamber - A fluid dynamic investigation}},}\
  }\href {\doibase 10.1038/339613a0} {\bibfield  {journal} {\bibinfo  {journal}
  {Nature}\ }\textbf {\bibinfo {volume} {339}},\ \bibinfo {pages} {613--616}
  (\bibinfo {year} {1989})}\BibitemShut {NoStop}%
\bibitem [{\citenamefont {Davaille}\ and\ \citenamefont
  {Jaupart}(1993)}]{davaille1993thermal}%
  \BibitemOpen
  \bibfield  {author} {\bibinfo {author} {\bibfnamefont {Anne}\ \bibnamefont
  {Davaille}}\ and\ \bibinfo {author} {\bibfnamefont {Claude}\ \bibnamefont
  {Jaupart}},\ }\bibfield  {title} {\enquote {\bibinfo {title} {Thermal
  convection in lava lakes},}\ }\href {\doibase 10.1029/93GL02008} {\bibfield
  {journal} {\bibinfo  {journal} {Geophys. Res. Lett.}\ }\textbf {\bibinfo
  {volume} {20}},\ \bibinfo {pages} {1827--1830} (\bibinfo {year}
  {1993})}\BibitemShut {NoStop}%
\bibitem [{\citenamefont {Polashenski}\ \emph {et~al.}(2012)\citenamefont
  {Polashenski}, \citenamefont {Perovich},\ and\ \citenamefont
  {Courville}}]{Polashenski2012}%
  \BibitemOpen
  \bibfield  {author} {\bibinfo {author} {\bibfnamefont {Chris}\ \bibnamefont
  {Polashenski}}, \bibinfo {author} {\bibfnamefont {Donald~K.}\ \bibnamefont
  {Perovich}}, \ and\ \bibinfo {author} {\bibfnamefont {Zoe}\ \bibnamefont
  {Courville}},\ }\bibfield  {title} {\enquote {\bibinfo {title} {The
  mechanisms of sea ice melt pond formation and evolution},}\ }\href
  {http://dx.doi.org/10.1029/2011JC007231} {\bibfield  {journal} {\bibinfo
  {journal} {J. Geophys. Res.}\ }\textbf {\bibinfo {volume} {117}} (\bibinfo
  {year} {2012})}\BibitemShut {NoStop}%
\bibitem [{\citenamefont {Polashenski}\ \emph {et~al.}(2017)\citenamefont
  {Polashenski}, \citenamefont {Golden}, \citenamefont {Perovich},
  \citenamefont {Skyllingstad}, \citenamefont {Arnsten}, \citenamefont
  {Stwertka},\ and\ \citenamefont {Wright}}]{Polashenski2017}%
  \BibitemOpen
  \bibfield  {author} {\bibinfo {author} {\bibfnamefont {Chris}\ \bibnamefont
  {Polashenski}}, \bibinfo {author} {\bibfnamefont {Kenneth~M.}\ \bibnamefont
  {Golden}}, \bibinfo {author} {\bibfnamefont {Donald~K.}\ \bibnamefont
  {Perovich}}, \bibinfo {author} {\bibfnamefont {Eric}\ \bibnamefont
  {Skyllingstad}}, \bibinfo {author} {\bibfnamefont {Alexandra}\ \bibnamefont
  {Arnsten}}, \bibinfo {author} {\bibfnamefont {Carolyn}\ \bibnamefont
  {Stwertka}}, \ and\ \bibinfo {author} {\bibfnamefont {Nicholas}\ \bibnamefont
  {Wright}},\ }\bibfield  {title} {\enquote {\bibinfo {title} {Percolation
  blockage: A process that enables melt pond formation on first year arctic sea
  ice},}\ }\href {\doibase 10.1002/2016JC011994} {\bibfield  {journal}
  {\bibinfo  {journal} {J. Geophys. Res.: Oceans}\ }\textbf {\bibinfo {volume}
  {122}},\ \bibinfo {pages} {413--440} (\bibinfo {year} {2017})}\BibitemShut
  {NoStop}%
\bibitem [{\citenamefont
  {Chandrasekhar}(1961)}]{chandrasekhar2013hydrodynamic}%
  \BibitemOpen
  \bibfield  {author} {\bibinfo {author} {\bibfnamefont {Subrahmanyan}\
  \bibnamefont {Chandrasekhar}},\ }\href
  {https://www.amazon.com/Hydrodynamic-Hydromagnetic-Stability-Dover-Physics-ebook/dp/B00C59C7ZA?SubscriptionId=0JYN1NVW651KCA56C102&tag=techkie-20&linkCode=xm2&camp=2025&creative=165953&creativeASIN=B00C59C7ZA}
  {\emph {\bibinfo {title} {Hydrodynamic and Hydromagnetic Stability}}}\
  (\bibinfo  {publisher} {Dover Publications},\ \bibinfo {year}
  {1961})\BibitemShut {NoStop}%
\bibitem [{\citenamefont {Chill{\`a}}\ and\ \citenamefont
  {Schumacher}(2012)}]{Chilla}%
  \BibitemOpen
  \bibfield  {author} {\bibinfo {author} {\bibfnamefont {F.}~\bibnamefont
  {Chill{\`a}}}\ and\ \bibinfo {author} {\bibfnamefont {J.}~\bibnamefont
  {Schumacher}},\ }\bibfield  {title} {\enquote {\bibinfo {title} {New
  perspectives in turbulent {Rayleigh}-{B{\'e}nard} convection},}\ }\href
  {\doibase 10.1140/epje/i2012-12058-1} {\bibfield  {journal} {\bibinfo
  {journal} {Eur. Phys. J. E}\ }\textbf {\bibinfo {volume} {35}},\ \bibinfo
  {pages} {58} (\bibinfo {year} {2012})}\BibitemShut {NoStop}%
\bibitem [{\citenamefont {Ahlers}\ \emph {et~al.}(2009)\citenamefont {Ahlers},
  \citenamefont {Grossmann},\ and\ \citenamefont {Lohse}}]{ahlers2009heat}%
  \BibitemOpen
  \bibfield  {author} {\bibinfo {author} {\bibfnamefont {Guenter}\ \bibnamefont
  {Ahlers}}, \bibinfo {author} {\bibfnamefont {Siegfried}\ \bibnamefont
  {Grossmann}}, \ and\ \bibinfo {author} {\bibfnamefont {Detlef}\ \bibnamefont
  {Lohse}},\ }\bibfield  {title} {\enquote {\bibinfo {title} {Heat transfer and
  large scale dynamics in turbulent {Rayleigh}-{B\'enard} convection},}\ }\href
  {\doibase 10.1103/RevModPhys.81.503} {\bibfield  {journal} {\bibinfo
  {journal} {Rev. Mod. Phys.}\ }\textbf {\bibinfo {volume} {81}},\ \bibinfo
  {pages} {503--537} (\bibinfo {year} {2009})}\BibitemShut {NoStop}%
\bibitem [{\citenamefont {Taylor}\ and\ \citenamefont
  {Feltham}(2004)}]{taylor2004model}%
  \BibitemOpen
  \bibfield  {author} {\bibinfo {author} {\bibfnamefont {P.~D.}\ \bibnamefont
  {Taylor}}\ and\ \bibinfo {author} {\bibfnamefont {D.~L.}\ \bibnamefont
  {Feltham}},\ }\bibfield  {title} {\enquote {\bibinfo {title} {A model of melt
  pond evolution on sea ice},}\ }\href {\doibase 10.1029/2004JC002361}
  {\bibfield  {journal} {\bibinfo  {journal} {J. Geophys. Res.: Oceans}\
  }\textbf {\bibinfo {volume} {109}},\ \bibinfo {pages} {C12007} (\bibinfo
  {year} {2004})}\BibitemShut {NoStop}%
\bibitem [{\citenamefont {Deser}\ \emph {et~al.}(2000)\citenamefont {Deser},
  \citenamefont {Walsh},\ and\ \citenamefont {Timlin}}]{Deser2000}%
  \BibitemOpen
  \bibfield  {author} {\bibinfo {author} {\bibfnamefont {Clara}\ \bibnamefont
  {Deser}}, \bibinfo {author} {\bibfnamefont {John~E.}\ \bibnamefont {Walsh}},
  \ and\ \bibinfo {author} {\bibfnamefont {Michael~S.}\ \bibnamefont
  {Timlin}},\ }\bibfield  {title} {\enquote {\bibinfo {title} {Arctic sea ice
  variability in the context of recent atmospheric circulation trends},}\
  }\href {\doibase
  https://doi.org/10.1175/1520-0442(2000)013<0617:ASIVIT>2.0.CO;2} {\bibfield
  {journal} {\bibinfo  {journal} {J. Climate}\ }\textbf {\bibinfo {volume}
  {13}},\ \bibinfo {pages} {617--633} (\bibinfo {year} {2000})}\BibitemShut
  {NoStop}%
\bibitem [{\citenamefont {Flocco}\ and\ \citenamefont
  {Feltham}(2007)}]{flocco2007continuum}%
  \BibitemOpen
  \bibfield  {author} {\bibinfo {author} {\bibfnamefont {Daniela}\ \bibnamefont
  {Flocco}}\ and\ \bibinfo {author} {\bibfnamefont {Daniel~L.}\ \bibnamefont
  {Feltham}},\ }\bibfield  {title} {\enquote {\bibinfo {title} {A continuum
  model of melt pond evolution on {Arctic} sea ice},}\ }\href {\doibase
  10.1029/2006JC003836} {\bibfield  {journal} {\bibinfo  {journal} {J. Geophys.
  Res.: Oceans}\ }\textbf {\bibinfo {volume} {112}},\ \bibinfo {pages} {C08016}
  (\bibinfo {year} {2007})}\BibitemShut {NoStop}%
\bibitem [{\citenamefont {Scott}\ and\ \citenamefont
  {Feltham}(2010)}]{scott2010}%
  \BibitemOpen
  \bibfield  {author} {\bibinfo {author} {\bibfnamefont {F.}~\bibnamefont
  {Scott}}\ and\ \bibinfo {author} {\bibfnamefont {D.~L.}\ \bibnamefont
  {Feltham}},\ }\bibfield  {title} {\enquote {\bibinfo {title} {A model of the
  three-dimensional evolution of arctic melt ponds on first-year and multiyear
  sea ice},}\ }\href {http://dx.doi.org/10.1029/2010JC006156} {\bibfield
  {journal} {\bibinfo  {journal} {J. Geophys. Res.: Oceans}\ }\textbf {\bibinfo
  {volume} {115}} (\bibinfo {year} {2010})}\BibitemShut {NoStop}%
\bibitem [{\citenamefont {Vancoppenolle}\ \emph {et~al.}(2009)\citenamefont
  {Vancoppenolle}, \citenamefont {Fichefet}, \citenamefont {Goosse},
  \citenamefont {Bouillon}, \citenamefont {Madec},\ and\ \citenamefont
  {Maqueda}}]{vancoppenolle2009simulating}%
  \BibitemOpen
  \bibfield  {author} {\bibinfo {author} {\bibfnamefont {Martin}\ \bibnamefont
  {Vancoppenolle}}, \bibinfo {author} {\bibfnamefont {Thierry}\ \bibnamefont
  {Fichefet}}, \bibinfo {author} {\bibfnamefont {Hugues}\ \bibnamefont
  {Goosse}}, \bibinfo {author} {\bibfnamefont {Sylvain}\ \bibnamefont
  {Bouillon}}, \bibinfo {author} {\bibfnamefont {Gurvan}\ \bibnamefont
  {Madec}}, \ and\ \bibinfo {author} {\bibfnamefont {Miguel Angel~Morales}\
  \bibnamefont {Maqueda}},\ }\bibfield  {title} {\enquote {\bibinfo {title}
  {Simulating the mass balance and salinity of {Arctic} and {Antarctic} sea
  ice. 1. model description and validation},}\ }\href {\doibase
  https://doi.org/10.1016/j.ocemod.2008.10.005} {\bibfield  {journal} {\bibinfo
   {journal} {Ocean Modelling}\ }\textbf {\bibinfo {volume} {27}},\ \bibinfo
  {pages} {33 -- 53} (\bibinfo {year} {2009})}\BibitemShut {NoStop}%
\bibitem [{\citenamefont {Podgorny}\ and\ \citenamefont
  {Grenfell}(1996)}]{podgorny1996}%
  \BibitemOpen
  \bibfield  {author} {\bibinfo {author} {\bibfnamefont {Igor~A.}\ \bibnamefont
  {Podgorny}}\ and\ \bibinfo {author} {\bibfnamefont {Thomas~C.}\ \bibnamefont
  {Grenfell}},\ }\bibfield  {title} {\enquote {\bibinfo {title} {Partitioning
  of solar energy in melt ponds from measurements of pond albedo and depth},}\
  }\href {\doibase 10.1029/96JC02123} {\bibfield  {journal} {\bibinfo
  {journal} {J. Geophys. Res.: Oceans}\ }\textbf {\bibinfo {volume} {101}},\
  \bibinfo {pages} {22737--22748} (\bibinfo {year} {1996})}\BibitemShut
  {NoStop}%
\bibitem [{\citenamefont {Alexiades}(1992)}]{alexiades1992mathematical}%
  \BibitemOpen
  \bibfield  {author} {\bibinfo {author} {\bibfnamefont {Vasilios}\
  \bibnamefont {Alexiades}},\ }\href@noop {} {\emph {\bibinfo {title}
  {Mathematical modeling of melting and freezing processes}}}\ (\bibinfo
  {publisher} {CRC Press},\ \bibinfo {year} {1992})\BibitemShut {NoStop}%
\bibitem [{\citenamefont {Shyy}\ \emph {et~al.}(2012)\citenamefont {Shyy},
  \citenamefont {Udaykumar}, \citenamefont {Rao},\ and\ \citenamefont
  {Smith}}]{shyy}%
  \BibitemOpen
  \bibfield  {author} {\bibinfo {author} {\bibfnamefont {W.}~\bibnamefont
  {Shyy}}, \bibinfo {author} {\bibfnamefont {H.S.}\ \bibnamefont {Udaykumar}},
  \bibinfo {author} {\bibfnamefont {M.M.}\ \bibnamefont {Rao}}, \ and\ \bibinfo
  {author} {\bibfnamefont {R.W.}\ \bibnamefont {Smith}},\ }\href
  {https://books.google.fr/books?id=vy4dccp5CnIC} {\emph {\bibinfo {title}
  {Computational Fluid Dynamics with Moving Boundaries}}},\ Dover Books\
  (\bibinfo  {publisher} {Dover Publications},\ \bibinfo {year}
  {2012})\BibitemShut {NoStop}%
\bibitem [{\citenamefont {Ulvrov\'a}\ \emph {et~al.}(2012)\citenamefont
  {Ulvrov\'a}, \citenamefont {Labrosse}, \citenamefont {Coltice}, \citenamefont
  {R\r{a}back},\ and\ \citenamefont {Tackley}}]{Ulvrova}%
  \BibitemOpen
  \bibfield  {author} {\bibinfo {author} {\bibfnamefont {M.}~\bibnamefont
  {Ulvrov\'a}}, \bibinfo {author} {\bibfnamefont {S.}~\bibnamefont {Labrosse}},
  \bibinfo {author} {\bibfnamefont {N.}~\bibnamefont {Coltice}}, \bibinfo
  {author} {\bibfnamefont {P.}~\bibnamefont {R\r{a}back}}, \ and\ \bibinfo
  {author} {\bibfnamefont {P.J.}\ \bibnamefont {Tackley}},\ }\bibfield  {title}
  {\enquote {\bibinfo {title} {Numerical modelling of convection interacting
  with a melting and solidification front: {Application} to the thermal
  evolution of the basal magma ocean},}\ }\href {\doibase
  https://doi.org/10.1016/j.pepi.2012.06.008} {\bibfield  {journal} {\bibinfo
  {journal} {Phys. Earth Planet. Inter.}\ }\textbf {\bibinfo {volume} {206}},\
  \bibinfo {pages} {51 -- 66} (\bibinfo {year} {2012})}\BibitemShut {NoStop}%
\bibitem [{\citenamefont {Malkus}(1954)}]{malkus_1954}%
  \BibitemOpen
  \bibfield  {author} {\bibinfo {author} {\bibfnamefont {W.~V.~R.}\
  \bibnamefont {Malkus}},\ }\bibfield  {title} {\enquote {\bibinfo {title} {The
  heat transport and spectrum of thermal turbulence},}\ }\href {\doibase
  10.1098/rspa.1954.0197} {\bibfield  {journal} {\bibinfo  {journal} {Proc.
  Royal Soc. A}\ }\textbf {\bibinfo {volume} {225}},\ \bibinfo {pages} {1161}
  (\bibinfo {year} {1954})}\BibitemShut {NoStop}%
\bibitem [{\citenamefont {Succi}(2001)}]{succi2001lattice}%
  \BibitemOpen
  \bibfield  {author} {\bibinfo {author} {\bibfnamefont {Sauro}\ \bibnamefont
  {Succi}},\ }\href@noop {} {\emph {\bibinfo {title} {The lattice {Boltzmann}
  equation: for fluid dynamics and beyond}}}\ (\bibinfo  {publisher} {Oxford
  university press},\ \bibinfo {year} {2001})\BibitemShut {NoStop}%
\bibitem [{\citenamefont {Huber}\ \emph {et~al.}(2008)\citenamefont {Huber},
  \citenamefont {Parmigiani}, \citenamefont {Chopard}, \citenamefont {Manga},\
  and\ \citenamefont {Bachmann}}]{Huber}%
  \BibitemOpen
  \bibfield  {author} {\bibinfo {author} {\bibfnamefont {Christian}\
  \bibnamefont {Huber}}, \bibinfo {author} {\bibfnamefont {Andrea}\
  \bibnamefont {Parmigiani}}, \bibinfo {author} {\bibfnamefont {Bastien}\
  \bibnamefont {Chopard}}, \bibinfo {author} {\bibfnamefont {Michael}\
  \bibnamefont {Manga}}, \ and\ \bibinfo {author} {\bibfnamefont {Olivier}\
  \bibnamefont {Bachmann}},\ }\bibfield  {title} {\enquote {\bibinfo {title}
  {{Lattice Boltzmann} model for melting with natural convection},}\ }\href
  {\doibase https://doi.org/10.1016/j.ijheatfluidflow.2008.05.002} {\bibfield
  {journal} {\bibinfo  {journal} {Int. J. Heat Fluid Fl.}\ }\textbf {\bibinfo
  {volume} {29}},\ \bibinfo {pages} {1469 -- 1480} (\bibinfo {year}
  {2008})}\BibitemShut {NoStop}%
\bibitem [{\citenamefont {Skyllingstad}\ \emph {et~al.}(2009)\citenamefont
  {Skyllingstad}, \citenamefont {Paulson},\ and\ \citenamefont
  {Perovich}}]{Skyllingstad}%
  \BibitemOpen
  \bibfield  {author} {\bibinfo {author} {\bibfnamefont {Eric~D.}\ \bibnamefont
  {Skyllingstad}}, \bibinfo {author} {\bibfnamefont {Clayton~A.}\ \bibnamefont
  {Paulson}}, \ and\ \bibinfo {author} {\bibfnamefont {Donald~K.}\ \bibnamefont
  {Perovich}},\ }\bibfield  {title} {\enquote {\bibinfo {title} {Simulation of
  melt pond evolution on level ice},}\ }\href {\doibase 10.1029/2009JC005363}
  {\bibfield  {journal} {\bibinfo  {journal} {J. Geophys. Res.: Oceans}\
  }\textbf {\bibinfo {volume} {114}},\ \bibinfo {pages} {C12019} (\bibinfo
  {year} {2009})}\BibitemShut {NoStop}%
\bibitem [{\citenamefont {Kim}\ \emph {et~al.}(2008)\citenamefont {Kim},
  \citenamefont {Lee},\ and\ \citenamefont {Choi}}]{Kim2008}%
  \BibitemOpen
  \bibfield  {author} {\bibinfo {author} {\bibfnamefont {Min~Chan}\
  \bibnamefont {Kim}}, \bibinfo {author} {\bibfnamefont {Dong~Won}\
  \bibnamefont {Lee}}, \ and\ \bibinfo {author} {\bibfnamefont {Chang~Kyun}\
  \bibnamefont {Choi}},\ }\bibfield  {title} {\enquote {\bibinfo {title} {Onset
  of buoyancy-driven convection in melting from below},}\ }\href {\doibase
  10.1007/s11814-008-0205-0} {\bibfield  {journal} {\bibinfo  {journal} {Korean
  J. Chem. Eng.}\ }\textbf {\bibinfo {volume} {25}},\ \bibinfo {pages}
  {1239--1244} (\bibinfo {year} {2008})}\BibitemShut {NoStop}%
\bibitem [{\citenamefont {Vasil}\ and\ \citenamefont
  {Proctor}(2011)}]{vasil_proctor_2011}%
  \BibitemOpen
  \bibfield  {author} {\bibinfo {author} {\bibfnamefont {G.~M.}\ \bibnamefont
  {Vasil}}\ and\ \bibinfo {author} {\bibfnamefont {M.~R.~E.}\ \bibnamefont
  {Proctor}},\ }\bibfield  {title} {\enquote {\bibinfo {title} {Dynamic
  bifurcations and pattern formation in melting-boundary convection},}\ }\href
  {\doibase 10.1017/jfm.2011.284} {\bibfield  {journal} {\bibinfo  {journal}
  {J. Fluid Mech.}\ }\textbf {\bibinfo {volume} {686}},\ \bibinfo {pages}
  {77--108} (\bibinfo {year} {2011})}\BibitemShut {NoStop}%
\bibitem [{SM()}]{SM}%
  \BibitemOpen
  \href@noop {} {\ }\bibinfo {note} {See Supplemental Material at [URL will be
  inserted by publisher] for videos of the results shown in Figs. 2 and
  7.}\BibitemShut {Stop}%
\bibitem [{\citenamefont {van~der Poel}\ \emph {et~al.}(2013)\citenamefont
  {van~der Poel}, \citenamefont {Stevens},\ and\ \citenamefont
  {Lohse}}]{van2015comparison}%
  \BibitemOpen
  \bibfield  {author} {\bibinfo {author} {\bibfnamefont {Erwin~P.}\
  \bibnamefont {van~der Poel}}, \bibinfo {author} {\bibfnamefont {Richard J.
  A.~M.}\ \bibnamefont {Stevens}}, \ and\ \bibinfo {author} {\bibfnamefont
  {Detlef}\ \bibnamefont {Lohse}},\ }\bibfield  {title} {\enquote {\bibinfo
  {title} {Comparison between two- and three-dimensional {Rayleigh}-{B\'enard}
  convection},}\ }\href {\doibase 10.1017/jfm.2013.488} {\bibfield  {journal}
  {\bibinfo  {journal} {J. Fluid Mech.}\ }\textbf {\bibinfo {volume} {736}},\
  \bibinfo {pages} {177--194} (\bibinfo {year} {2013})}\BibitemShut {NoStop}%
\bibitem [{\citenamefont {Grossmann}\ and\ \citenamefont
  {Lohse}(2000)}]{grossmann_lohse_2000}%
  \BibitemOpen
  \bibfield  {author} {\bibinfo {author} {\bibfnamefont {Siegfried}\
  \bibnamefont {Grossmann}}\ and\ \bibinfo {author} {\bibfnamefont {Detlef}\
  \bibnamefont {Lohse}},\ }\bibfield  {title} {\enquote {\bibinfo {title}
  {Scaling in thermal convection: a unifying theory},}\ }\href {\doibase
  10.1017/S0022112099007545} {\bibfield  {journal} {\bibinfo  {journal} {J.
  Fluid Mech.}\ }\textbf {\bibinfo {volume} {407}},\ \bibinfo {pages} {27--56}
  (\bibinfo {year} {2000})}\BibitemShut {NoStop}%
\bibitem [{\citenamefont {Stevens}\ \emph {et~al.}(2013)\citenamefont
  {Stevens}, \citenamefont {van~der Poel}, \citenamefont {Grossmann},\ and\
  \citenamefont {Lohse}}]{stevens2013unifying}%
  \BibitemOpen
  \bibfield  {author} {\bibinfo {author} {\bibfnamefont {Richard J. A.~M.}\
  \bibnamefont {Stevens}}, \bibinfo {author} {\bibfnamefont {Erwin~P.}\
  \bibnamefont {van~der Poel}}, \bibinfo {author} {\bibfnamefont {Siegfried}\
  \bibnamefont {Grossmann}}, \ and\ \bibinfo {author} {\bibfnamefont {Detlef}\
  \bibnamefont {Lohse}},\ }\bibfield  {title} {\enquote {\bibinfo {title} {The
  unifying theory of scaling in thermal convection: the updated prefactors},}\
  }\href {\doibase 10.1017/jfm.2013.298} {\bibfield  {journal} {\bibinfo
  {journal} {J. Fluid Mech.}\ }\textbf {\bibinfo {volume} {730}},\ \bibinfo
  {pages} {295?308} (\bibinfo {year} {2013})}\BibitemShut {NoStop}%
\bibitem [{\citenamefont {Davis}\ \emph {et~al.}(1984)\citenamefont {Davis},
  \citenamefont {M{\"u}ller},\ and\ \citenamefont
  {Dietsche}}]{davis1984pattern}%
  \BibitemOpen
  \bibfield  {author} {\bibinfo {author} {\bibfnamefont {S.~H.}\ \bibnamefont
  {Davis}}, \bibinfo {author} {\bibfnamefont {U.}~\bibnamefont {M{\"u}ller}}, \
  and\ \bibinfo {author} {\bibfnamefont {C.}~\bibnamefont {Dietsche}},\
  }\bibfield  {title} {\enquote {\bibinfo {title} {Pattern selection in
  single-component systems coupling {B\'enard} convection and
  solidification},}\ }\href {\doibase 10.1017/S0022112084001543} {\bibfield
  {journal} {\bibinfo  {journal} {J. Fluid Mech.}\ }\textbf {\bibinfo {volume}
  {144}},\ \bibinfo {pages} {133--151} (\bibinfo {year} {1984})}\BibitemShut
  {NoStop}%
\bibitem [{\citenamefont {Sugawara}\ \emph {et~al.}(2006)\citenamefont
  {Sugawara}, \citenamefont {Tamura}, \citenamefont {Satoh}, \citenamefont
  {Komatsu}, \citenamefont {Tago},\ and\ \citenamefont
  {Beer}}]{sugawara2007visual}%
  \BibitemOpen
  \bibfield  {author} {\bibinfo {author} {\bibfnamefont {M.}~\bibnamefont
  {Sugawara}}, \bibinfo {author} {\bibfnamefont {E.}~\bibnamefont {Tamura}},
  \bibinfo {author} {\bibfnamefont {Y.}~\bibnamefont {Satoh}}, \bibinfo
  {author} {\bibfnamefont {Y.}~\bibnamefont {Komatsu}}, \bibinfo {author}
  {\bibfnamefont {M.}~\bibnamefont {Tago}}, \ and\ \bibinfo {author}
  {\bibfnamefont {H.}~\bibnamefont {Beer}},\ }\bibfield  {title} {\enquote
  {\bibinfo {title} {Visual observations of flow structure and melting front
  morphology in horizontal ice plate melting from above into a mixture},}\
  }\href {\doibase 10.1007/s00231-006-0175-x} {\bibfield  {journal} {\bibinfo
  {journal} {Heat Mass Transf.}\ }\textbf {\bibinfo {volume} {43}},\ \bibinfo
  {pages} {1009} (\bibinfo {year} {2006})}\BibitemShut {NoStop}%
\bibitem [{\citenamefont {Hill}(1996)}]{hill1996}%
  \BibitemOpen
  \bibfield  {author} {\bibinfo {author} {\bibfnamefont {X.}~\bibnamefont
  {Hill}},\ }\emph {\bibinfo {title} {La convection sous un solide.
  Applications aux plan\`etes}},\ \href@noop {} {Ph.D. thesis},\ \bibinfo
  {school} {Universit\'e Paris 7} (\bibinfo {year} {1996}),\ \bibinfo {note}
  {th\`ese de doctorat, Institut de Physique du Globe de Paris}\BibitemShut
  {NoStop}%
\bibitem [{\citenamefont {Ulvrov{\'a}}(2012)}]{ulvrova2012dynamics}%
  \BibitemOpen
  \bibfield  {author} {\bibinfo {author} {\bibfnamefont {Martina}\ \bibnamefont
  {Ulvrov{\'a}}},\ }\emph {\bibinfo {title} {Dynamics of fluids and transport
  applied to the early Earth}},\ \href@noop {} {Ph.D. thesis},\ \bibinfo
  {school} {Ecole {N}ormale {S}up{\'e}rieure de {L}yon,ENS-LYON} (\bibinfo
  {year} {2012})\BibitemShut {NoStop}%
\bibitem [{\citenamefont {Shen}\ \emph {et~al.}(1996)\citenamefont {Shen},
  \citenamefont {Tong},\ and\ \citenamefont {Xia}}]{Shen96}%
  \BibitemOpen
  \bibfield  {author} {\bibinfo {author} {\bibfnamefont {Y.}~\bibnamefont
  {Shen}}, \bibinfo {author} {\bibfnamefont {P.}~\bibnamefont {Tong}}, \ and\
  \bibinfo {author} {\bibfnamefont {K.-Q.}\ \bibnamefont {Xia}},\ }\bibfield
  {title} {\enquote {\bibinfo {title} {Turbulent convection over rough
  surfaces},}\ }\href {\doibase 10.1103/PhysRevLett.76.908} {\bibfield
  {journal} {\bibinfo  {journal} {Phys. Rev. Lett.}\ }\textbf {\bibinfo
  {volume} {76}},\ \bibinfo {pages} {908--911} (\bibinfo {year}
  {1996})}\BibitemShut {NoStop}%
\bibitem [{\citenamefont {Ciliberto}\ and\ \citenamefont
  {Laroche}(1999)}]{Ciliberto99}%
  \BibitemOpen
  \bibfield  {author} {\bibinfo {author} {\bibfnamefont {S.}~\bibnamefont
  {Ciliberto}}\ and\ \bibinfo {author} {\bibfnamefont {C.}~\bibnamefont
  {Laroche}},\ }\bibfield  {title} {\enquote {\bibinfo {title} {Random
  roughness of boundary increases the turbulent convection scaling exponent},}\
  }\href {\doibase 10.1103/PhysRevLett.82.3998} {\bibfield  {journal} {\bibinfo
   {journal} {Phys. Rev. Lett.}\ }\textbf {\bibinfo {volume} {82}},\ \bibinfo
  {pages} {3998--4001} (\bibinfo {year} {1999})}\BibitemShut {NoStop}%
\bibitem [{\citenamefont {Stringano}\ \emph {et~al.}(2006)\citenamefont
  {Stringano}, \citenamefont {Pascazio},\ and\ \citenamefont
  {Verzicco}}]{Stringano2006}%
  \BibitemOpen
  \bibfield  {author} {\bibinfo {author} {\bibfnamefont {G.}~\bibnamefont
  {Stringano}}, \bibinfo {author} {\bibfnamefont {G.}~\bibnamefont {Pascazio}},
  \ and\ \bibinfo {author} {\bibfnamefont {R.}~\bibnamefont {Verzicco}},\
  }\bibfield  {title} {\enquote {\bibinfo {title} {Turbulent thermal convection
  over grooved plates},}\ }\href {\doibase 10.1017/S0022112006009785}
  {\bibfield  {journal} {\bibinfo  {journal} {J. Fluid Mech.}\ }\textbf
  {\bibinfo {volume} {557}},\ \bibinfo {pages} {307?336} (\bibinfo {year}
  {2006})}\BibitemShut {NoStop}%
\bibitem [{\citenamefont {Toppaladoddi}\ \emph {et~al.}(2015)\citenamefont
  {Toppaladoddi}, \citenamefont {Succi},\ and\ \citenamefont
  {Wettlaufer}}]{toppaladoddi2015tailoring}%
  \BibitemOpen
  \bibfield  {author} {\bibinfo {author} {\bibfnamefont {Srikanth}\
  \bibnamefont {Toppaladoddi}}, \bibinfo {author} {\bibfnamefont {Sauro}\
  \bibnamefont {Succi}}, \ and\ \bibinfo {author} {\bibfnamefont {John~S.}\
  \bibnamefont {Wettlaufer}},\ }\bibfield  {title} {\enquote {\bibinfo {title}
  {Tailoring boundary geometry to optimize heat transport in turbulent
  convection},}\ }\href {http://stacks.iop.org/0295-5075/111/i=4/a=44005}
  {\bibfield  {journal} {\bibinfo  {journal} {Europhys. Lett.}\ }\textbf
  {\bibinfo {volume} {111}},\ \bibinfo {pages} {44005} (\bibinfo {year}
  {2015})}\BibitemShut {NoStop}%
\bibitem [{\citenamefont {Zhu}\ \emph {et~al.}(2017)\citenamefont {Zhu},
  \citenamefont {Stevens}, \citenamefont {Verzicco},\ and\ \citenamefont
  {Lohse}}]{zhuPRL2017}%
  \BibitemOpen
  \bibfield  {author} {\bibinfo {author} {\bibfnamefont {Xiaojue}\ \bibnamefont
  {Zhu}}, \bibinfo {author} {\bibfnamefont {Richard J. A.~M.}\ \bibnamefont
  {Stevens}}, \bibinfo {author} {\bibfnamefont {Roberto}\ \bibnamefont
  {Verzicco}}, \ and\ \bibinfo {author} {\bibfnamefont {Detlef}\ \bibnamefont
  {Lohse}},\ }\bibfield  {title} {\enquote {\bibinfo {title}
  {Roughness-facilitated local 1/2 scaling does not imply the onset of the
  ultimate regime of thermal convection},}\ }\href@noop {} {\bibfield
  {journal} {\bibinfo  {journal} {Phys. Rev. Lett.}\ }\textbf {\bibinfo
  {volume} {119}},\ \bibinfo {pages} {064501} (\bibinfo {year}
  {2017})}\BibitemShut {NoStop}%
\bibitem [{\citenamefont {Bertrand}\ \emph {et~al.}(1999)\citenamefont
  {Bertrand}, \citenamefont {Binet}, \citenamefont {Combeau}, \citenamefont
  {Couturier}, \citenamefont {Delannoy}, \citenamefont {Gobin}, \citenamefont
  {Lacroix}, \citenamefont {Qu\'er\'e}, \citenamefont {M\'edale}, \citenamefont
  {Mencinger}, \citenamefont {Sadat},\ and\ \citenamefont
  {Vieira}}]{Bertrand1995}%
  \BibitemOpen
  \bibfield  {author} {\bibinfo {author} {\bibfnamefont {Olivier}\ \bibnamefont
  {Bertrand}}, \bibinfo {author} {\bibfnamefont {Bruno}\ \bibnamefont {Binet}},
  \bibinfo {author} {\bibfnamefont {Herv\'e}\ \bibnamefont {Combeau}}, \bibinfo
  {author} {\bibfnamefont {St\'ephane}\ \bibnamefont {Couturier}}, \bibinfo
  {author} {\bibfnamefont {Yves}\ \bibnamefont {Delannoy}}, \bibinfo {author}
  {\bibfnamefont {Dominique}\ \bibnamefont {Gobin}}, \bibinfo {author}
  {\bibfnamefont {Marcel}\ \bibnamefont {Lacroix}}, \bibinfo {author}
  {\bibfnamefont {Patrick~Le}\ \bibnamefont {Qu\'er\'e}}, \bibinfo {author}
  {\bibfnamefont {Marc}\ \bibnamefont {M\'edale}}, \bibinfo {author}
  {\bibfnamefont {Jure}\ \bibnamefont {Mencinger}}, \bibinfo {author}
  {\bibfnamefont {Hamou}\ \bibnamefont {Sadat}}, \ and\ \bibinfo {author}
  {\bibfnamefont {Gisele}\ \bibnamefont {Vieira}},\ }\bibfield  {title}
  {\enquote {\bibinfo {title} {Melting driven by natural convection a
  comparison exercise: first results},}\ }\href {\doibase
  https://doi.org/10.1016/S0035-3159(99)80013-0} {\bibfield  {journal}
  {\bibinfo  {journal} {Int. J. Thermal Sci.}\ }\textbf {\bibinfo {volume}
  {38}},\ \bibinfo {pages} {5 -- 26} (\bibinfo {year} {1999})}\BibitemShut
  {NoStop}%
\bibitem [{\citenamefont {Miller}\ and\ \citenamefont
  {Succi}(2002)}]{miller2002lattice}%
  \BibitemOpen
  \bibfield  {author} {\bibinfo {author} {\bibfnamefont {W.}~\bibnamefont
  {Miller}}\ and\ \bibinfo {author} {\bibfnamefont {S.}~\bibnamefont {Succi}},\
  }\bibfield  {title} {\enquote {\bibinfo {title} {A lattice {B}oltzmann model
  for anisotropic crystal growth from melt},}\ }\href {\doibase
  10.1023/A:1014510704701} {\bibfield  {journal} {\bibinfo  {journal} {J. Stat.
  Phys.}\ }\textbf {\bibinfo {volume} {107}},\ \bibinfo {pages} {173--186}
  (\bibinfo {year} {2002})}\BibitemShut {NoStop}%
\bibitem [{\citenamefont {Rasin}\ \emph {et~al.}(2005)\citenamefont {Rasin},
  \citenamefont {Miller},\ and\ \citenamefont {Succi}}]{rasin2005phase}%
  \BibitemOpen
  \bibfield  {author} {\bibinfo {author} {\bibfnamefont {I.}~\bibnamefont
  {Rasin}}, \bibinfo {author} {\bibfnamefont {W.}~\bibnamefont {Miller}}, \
  and\ \bibinfo {author} {\bibfnamefont {S.}~\bibnamefont {Succi}},\ }\bibfield
   {title} {\enquote {\bibinfo {title} {Phase-field lattice kinetic scheme for
  the numerical simulation of dendritic growth},}\ }\href {\doibase
  10.1103/PhysRevE.72.066705} {\bibfield  {journal} {\bibinfo  {journal} {Phys.
  Rev. E}\ }\textbf {\bibinfo {volume} {72}},\ \bibinfo {pages} {066705}
  (\bibinfo {year} {2005})}\BibitemShut {NoStop}%
\bibitem [{\citenamefont {Medvedev}\ and\ \citenamefont
  {Kassner}(2005)}]{medvedev2005lattice}%
  \BibitemOpen
  \bibfield  {author} {\bibinfo {author} {\bibfnamefont {Dmitry}\ \bibnamefont
  {Medvedev}}\ and\ \bibinfo {author} {\bibfnamefont {Klaus}\ \bibnamefont
  {Kassner}},\ }\bibfield  {title} {\enquote {\bibinfo {title} {Lattice
  {Boltzmann} scheme for crystal growth in external flows},}\ }\href {\doibase
  10.1103/PhysRevE.72.056703} {\bibfield  {journal} {\bibinfo  {journal} {Phys.
  Rev. E}\ }\textbf {\bibinfo {volume} {72}},\ \bibinfo {pages} {056703}
  (\bibinfo {year} {2005})}\BibitemShut {NoStop}%
\bibitem [{\citenamefont {Wen-Shu~Jiaung}(2001)}]{jiaung2001lattice}%
  \BibitemOpen
  \bibfield  {author} {\bibinfo {author} {\bibfnamefont {Chun-Pao~Kuo}\
  \bibnamefont {Wen-Shu~Jiaung}, \bibfnamefont {Jeng-Rong~Ho}},\ }\bibfield
  {title} {\enquote {\bibinfo {title} {{Lattice Boltzmann} method for the heat
  conduction problem with phase change},}\ }\href {\doibase
  10.1080/10407790150503495} {\bibfield  {journal} {\bibinfo  {journal} {Numer.
  Heat Tr. B-Fund.}\ }\textbf {\bibinfo {volume} {39}},\ \bibinfo {pages}
  {167--187} (\bibinfo {year} {2001})}\BibitemShut {NoStop}%
\bibitem [{\citenamefont {Chatterjee}\ and\ \citenamefont
  {Chakraborty}(2005)}]{chatterjee2005enthalpy}%
  \BibitemOpen
  \bibfield  {author} {\bibinfo {author} {\bibfnamefont {Dipankar}\
  \bibnamefont {Chatterjee}}\ and\ \bibinfo {author} {\bibfnamefont {Suman}\
  \bibnamefont {Chakraborty}},\ }\bibfield  {title} {\enquote {\bibinfo {title}
  {An enthalpy-based lattice {Boltzmann} model for diffusion dominated
  solid-liquid phase transformation},}\ }\href {\doibase
  https://doi.org/10.1016/j.physleta.2005.04.080} {\bibfield  {journal}
  {\bibinfo  {journal} {Phys. Lett. A}\ }\textbf {\bibinfo {volume} {341}},\
  \bibinfo {pages} {320 -- 330} (\bibinfo {year} {2005})}\BibitemShut {NoStop}%
\end{thebibliography}

%

\end{document}